\documentclass[onecolumn]{openjournal}

\usepackage{xcolor}
\definecolor{skyblue}{rgb}{0.53, 0.81, 0.92}
\definecolor{deepskyblue}{rgb}{0.0, 0.75, 1.0}

\newcommand{\code}[1]{\texttt{#1}}

\newcommand{\tm}{\textit{Telescope\_Moving}}
\newcommand{\gs}{\textit{Ghost\_Scatter}}

\newcommand{\lsp}{\code{legacypipe}}
\usepackage{listings}
\usepackage[utf8]{inputenc}

\usepackage{tikz}
\usetikzlibrary{shapes.geometric, arrows}

\usepackage{orcidlink}

\usepackage{hyperref}
\hypersetup{
    colorlinks=true,
    citecolor=blue,
}

\usepackage{array}
\usepackage{multirow}
\newcolumntype{C}[1]{>{\centering\arraybackslash}m{#1}}

\begin{document}

\title{A Semi-Supervised Learning Method for \\
the Identification of Bad Exposures in Large Imaging Surveys}

\author{Yufeng Luo\orcidlink{0000-0002-4623-0683}$^{1,2,3,*}$}

% \correspondingauthor{Yufeng Luo}
\email{*Corresponding author: yluo4@uwyo.edu}
% \email{$^*$Corresponding author: yluo4@uwyo.edu}

\author{Adam D. Myers$^{1}$}

\author{Alex Drlica-Wagner\orcidlink{0000-0001-8251-933X}$^{4,5,6,7}$}

\author{Dario Dematties\orcidlink{0000-0002-8726-7837}$^{3}$}

\author{Salma Borchani$^{1}$}

\author{Francisco Valdes$^{8}$}

\author{Arjun Dey$^{8}$}

\author{David Schlegel$^{9}$}

\author{Rongpu Zhou$^{9}$}

\author{DESI Legacy Imaging Surveys Team}

\affiliation{$^1$Department of Physics and Astronomy,
University of Wyoming,
1000 E. University Ave., Laramie, WY 82071, USA}
\affiliation{$^2$School of Computing,
University of Wyoming,
1000 E. University Ave., Laramie, WY 82071, USA}
\affiliation{$^3$Mathematics and Computer Science Division,
Argonne National Laboratory,
Lemont, IL}
\affiliation{$^4$Fermi National Accelerator Laboratory, P.O.\ Box 500, Batavia, IL 60510, USA}
\affiliation{$^5$Department of Astronomy and Astrophysics, University of Chicago, Chicago, IL 60637, USA}
\affiliation{$^6$Kavli Institute for Cosmological Physics, University of Chicago, Chicago, IL 60637, USA}
\affiliation{$^7$NSF-Simons AI Institute for the Sky (SkAI), 172 E. Chestnut St., Chicago, IL 60611, USA}
\affiliation{$^8$NSF NOIRLab, 950 N Cherry Ave, Tucson, AZ 85719, USA}
\affiliation{$^9$Lawrence Berkeley National Laboratory, 1 Cyclotron Road, Berkeley, CA 94720, USA}

\begin{abstract}

As the data volume of astronomical imaging surveys rapidly increases, traditional methods for image anomaly detection, such as visual inspection by human experts, are becoming impractical.
We introduce a machine-learning-based approach to detect poor-quality exposures in large imaging surveys, with a focus on the DECam Legacy Survey (DECaLS) in regions of low extinction (i.e., $E(B-V)<0.04$).
Our semi-supervised pipeline integrates a vision transformer (ViT), trained via self-supervised learning (SSL), with a k-Nearest Neighbor (kNN) classifier.
We train and validate our pipeline using a small set of labeled exposures observed by surveys with the Dark Energy Camera (DECam). A clustering-space analysis of where our pipeline places images labeled in ``good'' and ``bad'' categories suggests that our approach can efficiently and accurately determine the quality of exposures.
Applied to new imaging being reduced for DECaLS Data Release 11, our pipeline identifies 780 problematic exposures, which we subsequently verify through visual inspection.
Being highly efficient and adaptable, our method offers a scalable solution for quality control in other large imaging surveys.

\end{abstract}

\section{Introduction} \label{sec:intro}

The Astro2020 decadal survey \citep{astro2020} highlighted the increasing importance of advanced pipelines for processing data from large astronomical surveys, and how scientific results will crucially depend on these pipelines. Identifying bad exposures is a key step in any pipeline to optimize the quality of images from sky surveys.
However, this step can often be laborious.
Traditionally, the bad exposures can be labeled in two ways. First, the observers at the telescope can record the exposure condition and quality during observation. Some issues, such as shaking due to strong winds, can be noted. 
Further, some common issues, such as hot pixels and cosmic rays, can be detected and fixed using algorithms and available software \citep{CRreject,astropy:2022}. Some other issues, like seeing and Point Spread Function (PSF) variations, are usually detectable during photometric reductions \citep{Tractor2016}. Beyond this, more
specific problems, such as ghosting and scattered light, have been historically difficult to handle and have often required special algorithmic treatment.
Although some solutions to these problems exist \citep{scatteredlight, osti_1690257}, they are usually built for a specific telescope or camera.
In this case, astronomers have to visually inspect exposures to determine which might be problematic and/or evaluate the image quality during the overall data reduction process.
Astronomers often also have to explicitly construct separate models or metrics to study each problematic imaging systematic.

As larger surveys proceed to everyday operations, the number of exposures in their associated imaging campaigns is likely to become too high for a human expert to visually inspect every image. Further, as the number of pixels achieved by cameras increases, features in images are likely to become more complex, which will make the identification of bad exposures even more taxing.
As an example, the LSSTCam on the Vera C.\ Rubin telescope utilizes a 3.2-gigapixel CCD detector and can observe 10{,}000 square degrees of sky in a single filter every three clear nights \citep{LSSTDesign}.
Despite the increase in scope and complexity of future imaging surveys, the quality of exposures in large campaigns will remain essential to downstream tasks, such as astrometric \citep[e.g.,][]{Lang_2009}, and photometric \citep[e.g.,][]{Stubbs_2006} measurements, and target selection to facilitate follow-up studies \citep[e.g.,][]{Myers_2023}. The bottom line is that unintentionally including bad exposures in the datasets produced by imaging surveys will always be detrimental to downstream scientific results.
Therefore, understanding and identifying problematic exposures will be a difficult but crucial element of next-generation imaging surveys. 

While large astronomical surveys have been expanding in volume and quality, more efficient data processing and analysis algorithms have already started to become desirable to take the burden off human inspectors. For example, \cite{Zhang_2020} and \cite{Chang_2021} used Convolutional Neural Networks (CNNs) to, respectively, identify cosmic rays in Hubble Space Telescope images, and detect ghosts and scattered light in Dark Energy Survey (DES) images;
\cite{deepGB} applied a method based on a Mask Region-Based Convolutional Neural Network (R-CNN) to predict the location of ghosts and scattered light in Dark Energy Camera \citep[henceforth DECam;][]{decam} exposures; and \cite{assessAutoML} applied existing image quality assessment (IQA) algorithms developed for general computer vision applications to score images drawn from the Dark Energy Spectroscopic Instrument (DESI) Legacy Imaging Surveys \citep[dubbed LS in the following text;][]{DESI_LS_Overview}, based on a predefined set of bad images.

%Recent advancements in machine learning algorithms and hardware provide the capacity for processing big data. From a model architecture perspective, the transformer architecture \citep{transfomerVaswani} has shown extraordinary performance not only on natural language processing tasks but also on image pattern recognition \citep{ViT2020}. On training algorithm, the Self-Supervised Learning (SSL) emerges as a useful technique for large dataset without labels.

In this work, we focus on using recently developed machine learning (henceforth ML) algorithms to identify bad exposures from DECam.
This is a complex problem as each DECam exposure comprises 62 individual CCDs (or 61 CCDs with N30 masked out\footnote{More details can be found at: \url{https://noirlab.edu/science/programs/ctio/instruments/Dark-Energy-Camera/Status-DECam-CCDs}}), and problems can be present both on large scales (i.e.,\ the entire exposure) and small scales (i.e.,\ individual CCDs, or even groups of pixels).
Some issues with the DECam exposures have been mitigated by the DECam Community Pipeline \citep[CP;][]{CP_pub}, which is a data reduction pipeline that processes DECam images and produces calibrated images and catalogs.
However, some issues, such as ghosting and scattered light, are not fully addressed by the CP, and thus require further investigation.
To address this complexity, our pipeline adopts an ML approach based on a Vision Transformer \citep[ViT;][]{ViT2020}.
ViT is a deep neural network that splits an image into small patches and learns patterns from them.
It is based on the transformer architecture \citep{transfomerVaswani} to capture complex patterns in images, often outperforming traditional convolutional neural networks (CNNs) when trained on large datasets. In particular, we use a pretrained existing ViT in the context of version 2 of the self-DIstillation with No Labels (DINOv2) framework \citep{caron2021emerging, oquab_dinov2_2023}, a self-supervised method that learns from images without human-provided labels. The original ViT was trained with ImageNet \citep{ImageNet}, and we apply that trained ViT to generate a numeric vector ``\textit{embedding}'' that summarizes the image's content. An embedding is essentially a representation of the features of an image in a high-dimensional parameter space.
We then use a k-Nearest Neighbor (kNN) algorithm to classify these embeddings after dimension reduction, and assign categories to images based on a labeled training set.

We aim to simultaneously detect multiple potential problems in an exposure, and use
a pipeline that only requires a relatively small training and testing dataset, typically comprising less than 5\% of the whole exposure dataset.
Because our model learns directly from the data, it can be adapted to other telescopes, although we choose to train our model using a dataset from DECam.
%We will discuss the details of the implementation in section \ref{sec:method}.
%To clarify the usage of synonyms, we refer to the full focal plane exposure image as \textbf{exposure} and to each single CCD image inside an exposure as \textbf{image}. 
Our pipeline is also relatively {\em fast} for inference. To tackle the large volume of data, our framework uses Graphics Processing Units (GPU). GPUs are a highly adaptable and efficient hardware for big datasets and are widely used in ML fields \citep{gpu_ML}. Moreover, we take advantage of multi-GPU parallel processing to achieve higher inference speed as long as the model can be fitted into each single GPU. 

This paper is organized as follows: Section \ref{sec:dataset} outlines the dataset we utilize, including the source and volume of the exposures, and the labels associated with some of the exposures.
In Section \ref{sec:method}, we discuss the details of our pipeline, the use of ViT and the layout of the classification task. We also discuss the training and validation procedure used by our pipeline.
In Section \ref{sec:results} we analyze the clustering and classification results from our model. We interpret our results, and compare our model with existing pipelines in Section \ref{sec:discussion}. Finally, we discuss limitations of our current framework and suggest future improvements in Sections \ref{sec:discussion} and \ref{sec:conclusion}.

\section{Dataset}\label{sec:dataset}

The exposures in our dataset are mostly drawn from the LS Data Releases 9, 10, and 11 which we henceforth refer to as DR9, DR10 and DR11.
The LS comprise three surveys: the Beijing-Arizona Sky Survey \citep[BASS;][]{BASS}, Mayall z-band Legacy Survey (MzLS), and DECam Legacy Survey \citep[DECaLS;][]{DESI_LS_Overview}.\footnote{We use LS to refer specifically to the DECaLS portion of the LS in the following text.}
We focus our analysis on DECam images from DECaLS collected with the Victor M. Blanco 4-meter Telescope\footnote{\url{https://noirlab.edu/public/programs/ctio/victor-blanco-4m-telescope/}} at the Cerro Tololo Inter-American Observatory (CTIO) in Chile.
DR11 also includes exposures from the DECam Local Volume Exploration Survey \citep[DELVE;][]{delve_dr1, delve_dr2}, which are incorporated in our dataset.
All of the images we consider in this paper are inputs to, or have been reduced by, the DECam Community Pipeline \citep[CP;][see Section~\ref{subsec:cpExtCut}]{CP_pub}.
A small fraction of the exposures are labeled by experts as problematic across a range of different categories. These labels come from three major sources: the DECam bad exposure list (see Section \ref{subsec:badExpList}), previous experts' visual inspection labels from DELVE, and a wider list of inspections flagged for the DES (Section \ref{subsec:expertLabel} and Appendix \ref{app:data}). In addition, we use bad exposure flags developed for DR9 and DR10 to narrow down the selection of good exposures (see Section \ref{subsec:legacypipeFlag}).
The numbers of categories that can be used as labels, and the numbers of bad exposures provided by each source are listed in Table \ref{tab:datasetCompose}.
There are 256,701 exposures in DR9 and DR10, and 18,315 exposures are labeled with at least 1 CCD image as bad by human experts, which is about 7\% of the total exposures.

Our task is to identify bad exposures from the unlabeled exposures based on the labeled exposures, which is a problem that is well-suited to a multi-class classification-task design. 
With this problem design, a balanced dataset is essential in the training and evaluation process for the classification. We therefore divide our dataset into three exclusive samples:
\begin{itemize}
    \item A training dataset based on 70\% of the labeled exposures and an equal amount of unlabeled good exposures. A 3-fold cross-validation \citep[CV;][]{Stone_1974} was used in the training.
    \item A testing dataset based on 30\% of the labeled exposures and an equal amount of unlabeled good exposures.
    \item An inference dataset that includes all exposures that are unlabeled excluding exposures used in the training and testing process. 
\end{itemize}

\begin{table}[h]
    \centering
    \begin{tabular}{|c|c|c|}
        \hline
        Source & Number of reported bad exposures & Number of categories \\
        \hline
        DECam bad exposure list & 3014 & 7 \\
        \hline
        DES list & 6667 & 61 \\
        \hline
        DELVE list & 8634 & 37 \\
        \hline
        LS pipeline (see \S\ref{subsec:legacypipeFlag}) flags & 119462 & 19 \\
        \hline
    \end{tabular}
    \caption{The number of bad exposures and categories from each source. The DES and DELVE samples are similar, except that DES includes additional ``niche'' categories. 
    %We only selected the most common categories from these two lists.
    }
    \label{tab:datasetCompose}
\end{table}

To balance the training, we added another 10{,}000 ``good'' exposures that are not included in any of the three samples listed above, although it is certainly possible that a small, but unknown, fraction of these ``good'' images include some exposures that have significant issues.
\cite{rolnick2018deeplearningrobustmassive} has shown that ML algorithms can be robust in classification tasks where a small fraction of label noise (i.e., mislabeled data) exists.

\subsection{DECaLS Bad Exposure List}\label{subsec:badExpList}

The DECaLS bad exposure list records 3{,}014 bad exposures identified during DECaLS observations and recorded in an observing log file named  \code{bad\_expid.txt}\footnote{\url{https://github.com/legacysurvey/legacypipe/blob/DR10.2/py/legacyzpts/data/decam-bad_expid.txt}}.
The issues recorded in the log file can be divided into three major categories: exposure quality, observing conditions, and instrument issues.
The log file uses three main flags, which are ``possibly saturated'' (denoted as 1 in \code{bad\_expid.txt}), ``seeing'' (denoted as 2), and ``other sources'' (denoted as 3). 
%We used the first two flags directly to indicate bad exposures.
%These issues span a large domain and each exposure may have slightly different severity.
As ``other sources'' denotes a wide range of issues of varying severity, we consider exposures in the list satisfying the following criteria as ``bad'':
\begin{itemize}
    \item the exposure flag is set to 1 or 2; or
    \item the exposure flag is set to 3, and either:
    \begin{itemize}
        \item \textit{expfactor} is included in the comment and is less than 1; or
        \item the ``other sources'' comment contains one of the keywords: \textit{telescope moving}, \textit{tracking}, \textit{PSF} or \textit{focus}.
    \end{itemize}
\end{itemize}

\noindent Here, \textit{expfactor} is an exposure factor defined as the ratio between the effective exposure time and the nominal exposure time. It is used to indicate the amount of light collected by the CCDs during the exposure after accounting for the transparency, seeing and instrument throughout. A value lower than 1 indicates that the exposure is not optimal.

%The ``other sources" flag usually has associated comments, which we parsed to capture keywords consistent with the above categories.
%We expect that seeing, transparency, and telescope instability will influence scales across an entire exposure, whereas other categories may be limited to smaller scales that are more consistent with a single CCD.

Throughout the rest of this paper, we will refer to this list of bad exposures as ``the DECaLS Bad Exposure List'' or simply ``the DECaLS list.''

\subsection{Expert-Labeled Exposures}\label{subsec:expertLabel}

The expert labels we use are provided by astronomers who visually inspected the exposures and evaluated their quality, in particular in the context of DELVE and DES.
%Both DELVE and DES are imaging campaigns that use DECam on the Blanco 4m telescope at Cerro Tololo Inter-American Observatory (CTIO).
The aim of DES \citep{DES2016} is to survey 5000 deg$^2$ of the southern sky in five optical bands ($grizY$).
The DES expert-labeled bad exposure list is based on the DES Year 6 (Y6) data release.
There are 122{,}531 CCD images labeled as bad from 6{,}667 exposures.
These exposures are categorized into 61 different types of comments, but we combined comments that signified similar issues.
%There are two main kinds of comments we combined. 
%The first type of comments we combined are associated with the same imaging patterns but are expressed slightly differently. 
For example, we incorporated both ``HeavyClouds'' and ``Clouds and transparency'' into the \textit{Clouds\_Transparency} category.
%, and we combined ``telescope moving'' and ``telescope tracking failure.'' %We are interested in the imaging pattern instead of the reasons, so we also combined them.

DELVE uses DECam to study ultra-faint galaxies around the Milky Way, Magellanic Clouds, and isolated dwarf galaxies in the Local Volume \citep{delve_dr1}.
The expert-labeled bad exposures we use in this paper were assembled on 2023 July 25 as an input into DELVE DR3 \citep{Tan_2025}.
The expert-labeled DELVE bad exposure dataset contains 51{,}445 problematic CCD images from 8{,}634 exposures.
There are 37 bad categories associated with these exposures.

The bad categories in DES do not perfectly match those in DELVE, so we created a unified list of bad categories to combine identified CCD images from the two surveys.
After combining similar cases, we chose the 10 most frequent issues appearing in the DES and/or DELVE datasets as our bad exposure labels.
The final number of images is shown in Table \ref{tab:category}.
These two datasets provide a large number of labeled bad exposures, which is an essential prerequisite for effectively training the classifier we use in this paper.
Throughout the rest of this paper, we will refer to the Expert-Labeled Exposures List as ``the human expert list''

\subsection{{\tt legacypipe} additional flags}\label{subsec:legacypipeFlag}

{\tt legacypipe}\footnote{code available at \url{https://github.com/legacysurvey/legacypipe}} is a pipeline based on the {\tt Tractor} \citep{Tractor2016} that is used to process the LS imaging data. As part of the reduction process, {\tt legacypipe} provides a set of flags to indicate the quality of each exposure \citep[see, e.g.][for more information]{DESI_LS_Overview}. The quality flags are stored in the survey CCD FITS file as integers that combine bits ranging from $2^{0}$ to $2^{18}$ where each individual bit indicates a specific condition.%, corresponding to 19 different potential issues.
\footnote{e.g.,\ \url{https://www.legacysurvey.org/dr10/bitmasks/\#ccd-cuts}}

We used the {\tt legacypipe} flags to further prune exposures 
%in the good exposure candidate sets 
that are not in the DECaLS list or the human expert list.
The total number of exposures with bad {\tt legacypipe} flags is 33.21\% of the total exposures in DR9 and DR10.
46.54\% of all exposures include at least one CCD with a bad {\tt legacypipe} flag.
Applying the {\tt legacypipe} flag cut removes these exposures and minimizes the contamination of the good exposure set by bad CCDs. 
We only selected exposures with no bad {\tt legacypipe} flags as good exposures in our training and testing datasets. This will improve the purity of the good exposure set, leading to a more robust training and testing process.

\subsection{Community Pipeline}\label{subsec:cpExtCut}

The DECam Community Pipeline \citep[CP;][]{CP_pub} is an imaging reduction pipeline that is used to process the DECaLS.
%The DECaLS images are processed by various CP versions, each of which targets some specific problem and, generally, later CP versions improve on earlier ones.
The range of CP versions used in the LS is from V4.10 through V5.5LS, with all processing performed later than 2019-06-06.\footnote{Information on CP versions: \url{https://desi.lbl.gov/trac/wiki/DecamLegacy/CPVersions}}
For each exposure, the CP produces many reduced data files and diagnostics. We chose to use the \code{ooi} images\footnote{For a more detailed description see \url{https://noirlab.edu/science/index.php/data-services/data-reduction-software/csdc-mso-pipelines/pl206\#tab10}} produced by the CP for our training process because those images are used for stacking and generating the small regions of the sky (``bricks'') that are defined to facilitate LS processing. These \code{ooi} images are better calibrated to the large-scale background than other LS products, and best represent the final LS science images.
Some exposures do not have associated \code{ooi} files because the CP processing ended prematurely due to problems in the exposures.
Since those exposure issues are usually very obvious, we ignored them during our training as they did not provide additional useful information.

\subsection{Extinction Cut}\label{subsec:extCut}

Because the original Key Projects of the DESI experiment \citep{desicollaboration2016desiexperimentisciencetargeting}, for which the LS was produced, focused on mainly extragalactic sources, the imaging catalog used for targeting also prioritizes sky areas with lower Galactic extinction to minimize the effect of dust attenuation.
High-extinction exposures usually have dust features that will produce patterns that interfere with the training of our ML model.
%, because the dust features in the exposures can lead to the detection of additional (unwanted) patterns.
Therefore, to minimize the influence of Galactic dust, we applied an extinction cut to exclude regions where $E(B-V) \ge 0.04$ \citep{sfd}, further down-selecting exposures for training.

\subsection{Categories of Bad Exposures}\label{subsec:pattern}

As noted in Section \ref{subsec:expertLabel}, we choose to reduce our dataset to the 11 categories of bad exposures that are most common and we find can best generalize typical problems.
The \textit{Saturated} category only appears in the DECaLS list, while the other 10 categories appear in both the human expert list and the DECaLS list.
Each of the categories corresponds to a specific issue for the images, and is drawn from the source list noted in Table \ref{tab:category}.
We provide a brief synopsis of each category below:

\begin{enumerate}
    \item \textit{Saturated}: This corresponds to a near-saturated exposure (generally taken near to twilight). The labeled data in this category is from the DECaLS list where the exposure flag is set to 1.
    \item \textit{Clouds\_transparency}: Cloud or humidity issues that affect the transparency during an exposure. The data is from the DECaLS list where the exposure flag is 2 or the comment contains either the keyword \textit{transparency} or \textit{clouds}.
    \item \textit{PSF}: The Point Spread Function issue. These sorts of issues can also typically be detected by standard (non-ML) photometry pipelines.
    \item \textit{Nonoptimal\_exp}: The exposure is underexposed or some similar exposure-related issue. This data is from the DECaLS list, where the keyword \texttt{expfactor} is included in a comment and is \textit{much} less than 1.
    \item \textit{Ghost\_scatter}: Ghosting or scattered light is present in an image. We follow the convention of the DES bad exposure list to combine these two effects into one category. These effects are usually caused by very bright sources within or near a given field.
    \item \textit{NObjects}: This label is provided by the DES Data Management processing pipeline \citep{Morganson_2018}. The category indicates that an anomalously large number of sources were detected in the image. This effect may be caused by issues with excess noise or difficulty in sky background estimation.
    \item \textit{Bad\_CCD}: The CCD image has no data or has an amplifier issue or has a hot spot (i.e., pixel-level defects).
    \item \textit{Noise}: The CCD includes some uncharacteristic electronic noise.% that is not fixed by dark or zero frames during data reduction. ADW: I don't see how darks or biases would "fix" electronic noise, so I've removed this statement.
    \item \textit{Wonky}: This category includes artifacts in exposures with patterns that do not belong to any other category and can be caused by miscellaneous reasons.
    \item \textit{Telescope\_moving}: The telescope tracking failed or the telescope unexpectedly moved.
    \item \textit{Out\_of\_focus}: The exposure was out of focus.
\end{enumerate}

\begin{figure}[h]
    \gridline{
        \fig{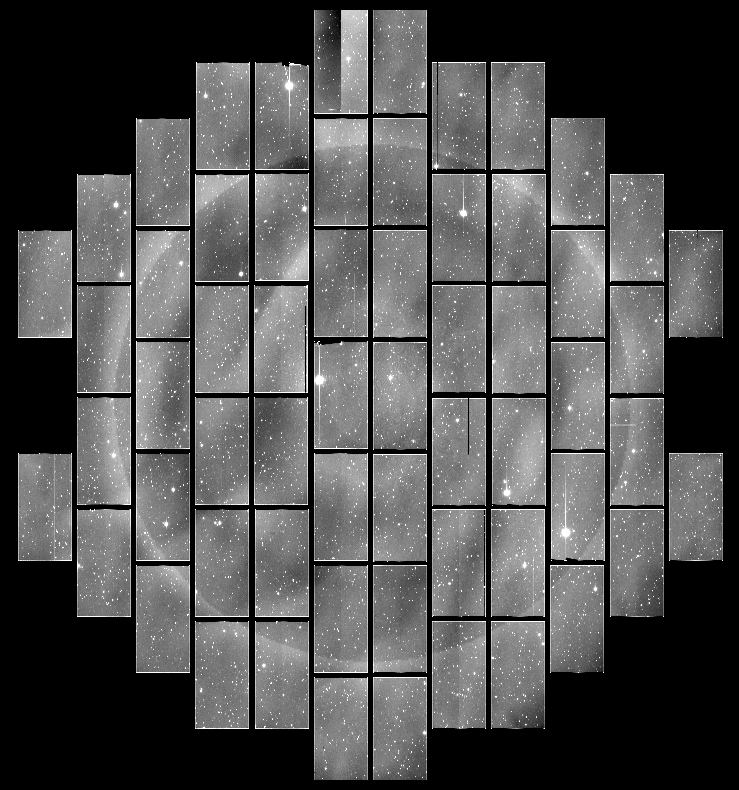}{0.3\textwidth}{(a) Saturated \newline expnum=347737}
        \fig{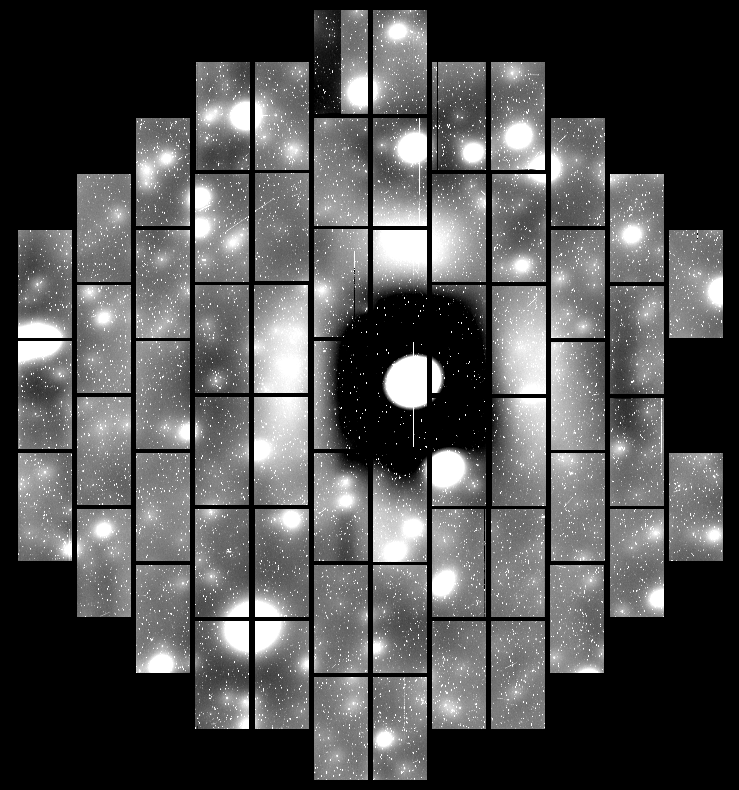}{0.3\textwidth}{(b) Transparency \newline expnum=212388}
        \fig{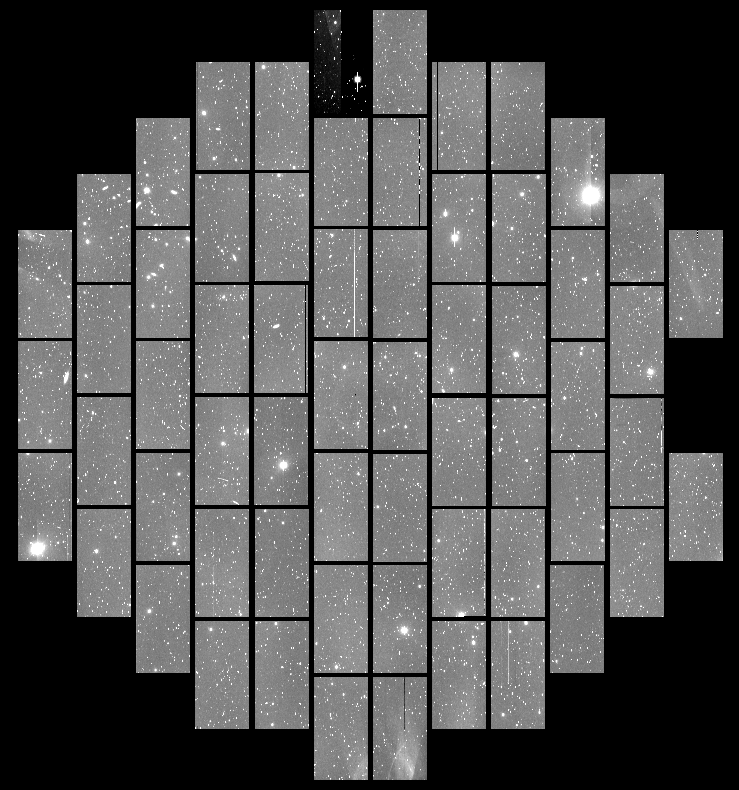}{0.3\textwidth}{(c) PSF \newline expnum=246860}
    }
    \gridline{
        \fig{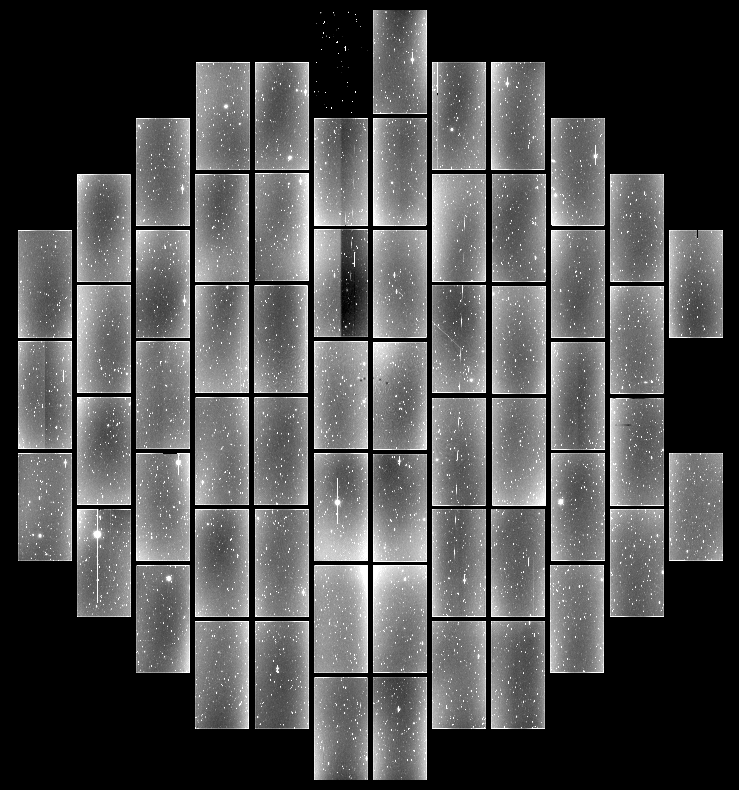}{0.3\textwidth}{(d) Non-optimal Exposure \newline expnum=663242}
        \fig{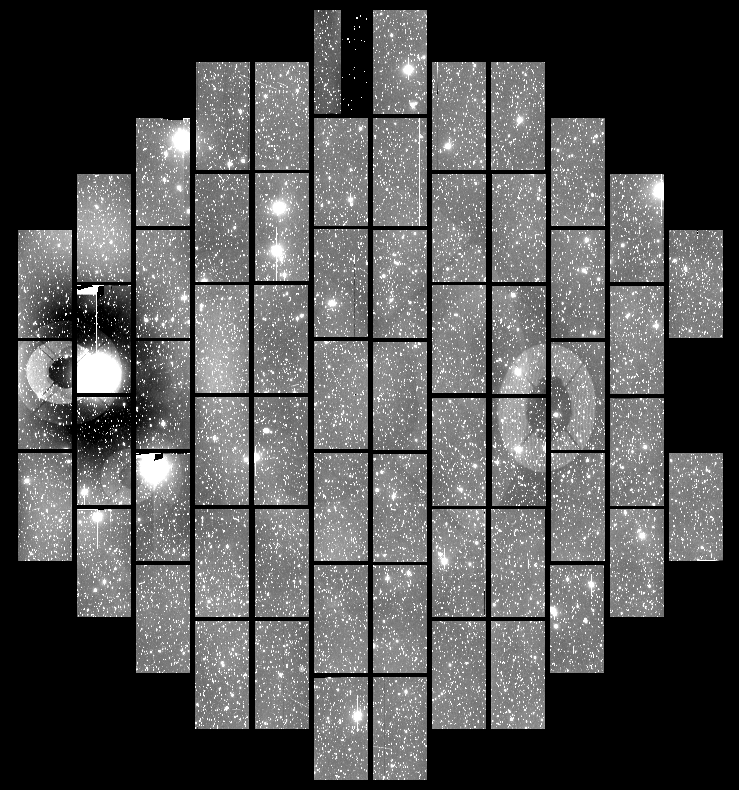}{0.3\textwidth}{(e) Ghosting \newline expnum=801945}
        \fig{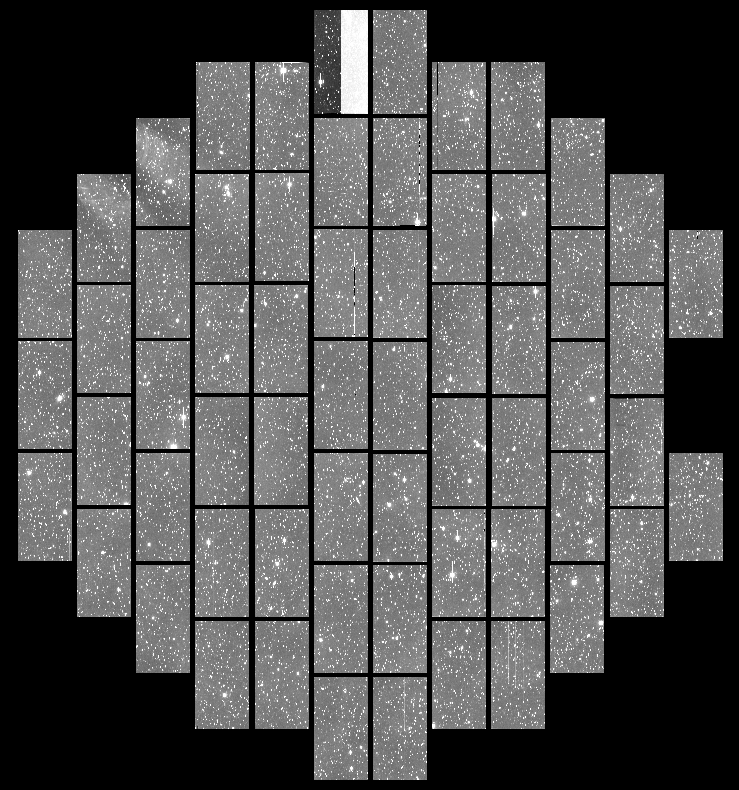}{0.3\textwidth}{(f) Scattered Light \newline expnum=198688}
    }
    \caption{A set of representative bad exposures in each category. These examples highlight why some common issues, like \textit{PSF} and \textit{NObjects} can be hard for visual inspectors to detect. Note that some of the exposures have been scaled to highlight the relevant feature. These sorts of scalings are frequently applied and might affect human experts' judgment, whereas the pipeline we describe in this paper uses raw exposures without any additional processing. It is worth noting that the saturated exposure displays a pattern similar to a flat fielding issue. This is because saturated exposures are mostly taken near twilight when the sky background flux-level will be very high. Therefore, the circular pattern that is visible in a saturated image typically corresponds to the instrument response after the flat-field correction.}
    \label{fig:example-bad-1}
\end{figure}

\begin{figure}
    \centering
    \gridline{
        \fig{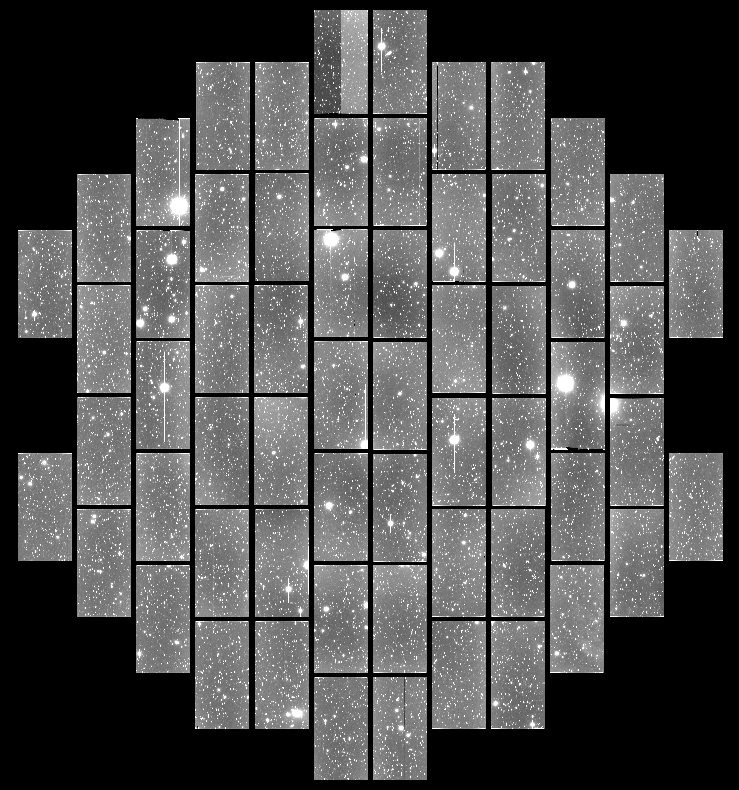}{0.3\textwidth}{(g) NObjects \newline expnum=563184}
        \fig{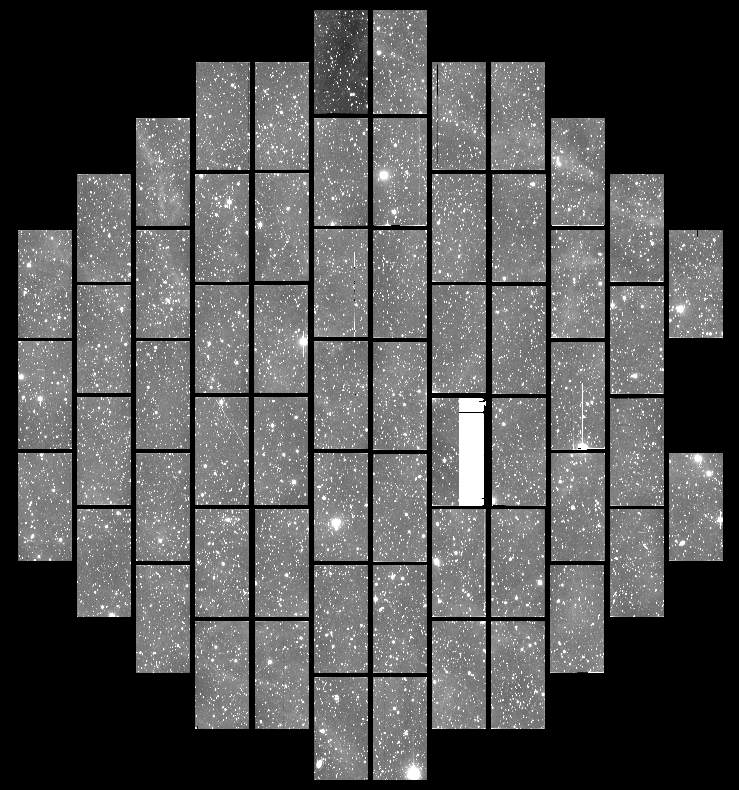}{0.3\textwidth}{(h) Bad CCD \newline expnum=767540}
        \fig{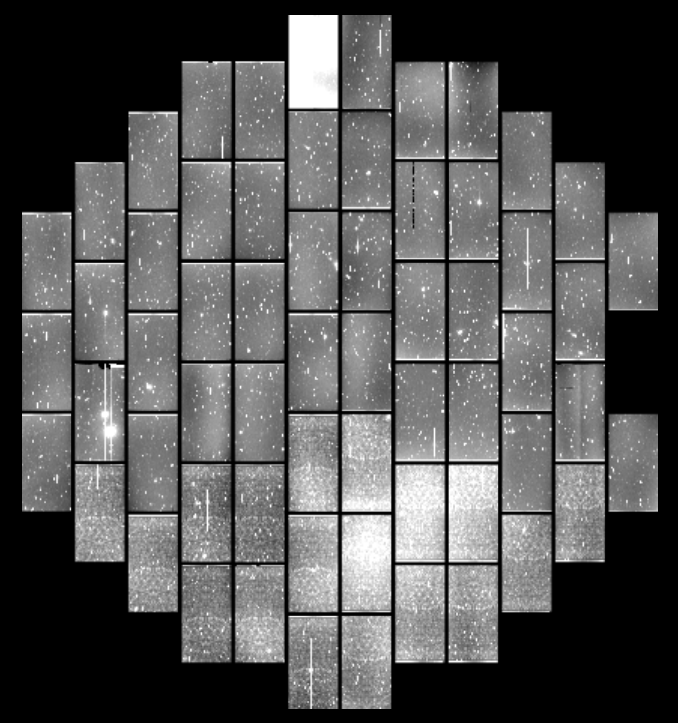}{0.3\textwidth}{(i) Noise \newline expnum=253706}
    }
    \gridline{
        \fig{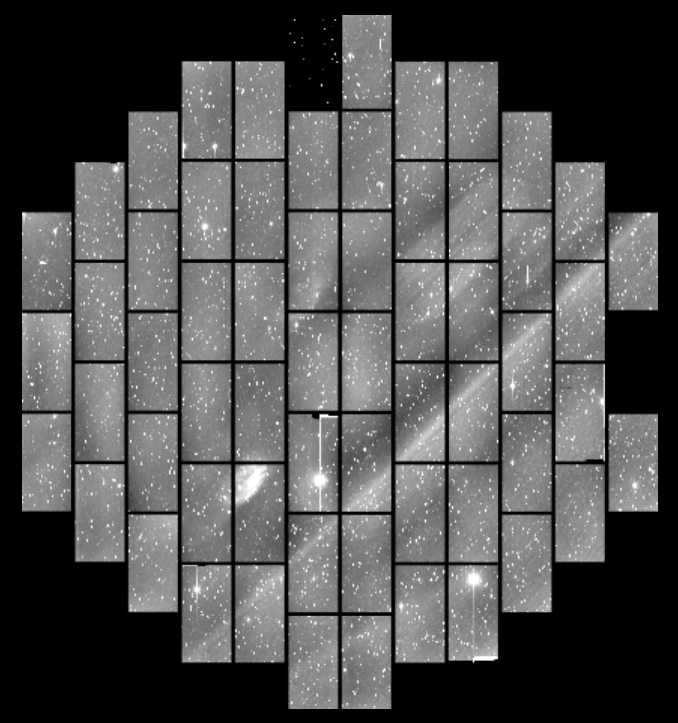}{0.3\textwidth}{(j) Wonky \newline expnum=989307}
        \fig{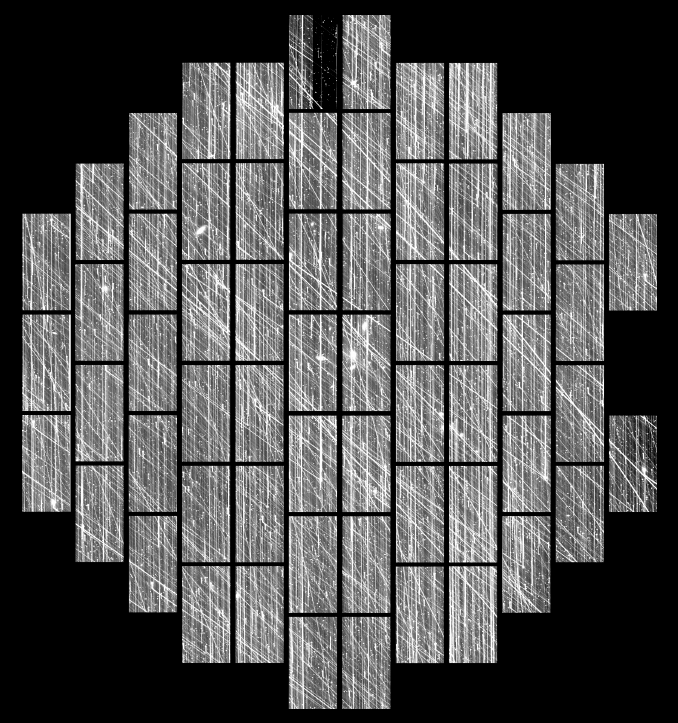}{0.3\textwidth}{(k) Telescope\_Moving\newline expnum=930303}
    }
    \caption{Continuation of Figure \ref{fig:example-bad-1}. This figure shows example exposures in the other 5 categories.}
    \label{fig:example-bad-2}
\end{figure}

As discussed in Sec.\ \ref{sec:intro}, some issues, such as \textit{PSF}, \textit{Out of focus}, and \textit{Saturated} can be detected by existing automated methods. We decided to retain these categories for two reasons.
First, these well-characterized issues can be used to validate our pipeline. %, as they are already known to be problematic and can serve as a benchmark for our pipeline performance.
Second, because the class labels our model considers are exclusive, including well-characterized issues allows the model to utilize \textit{complementary} information about less-well-characterized patterns,
which can improve the performance of the model across all  categories.

\subsection{Balancing the Dataset}

\begin{figure}[h]
    \centering
    \includegraphics[width=0.7\linewidth]{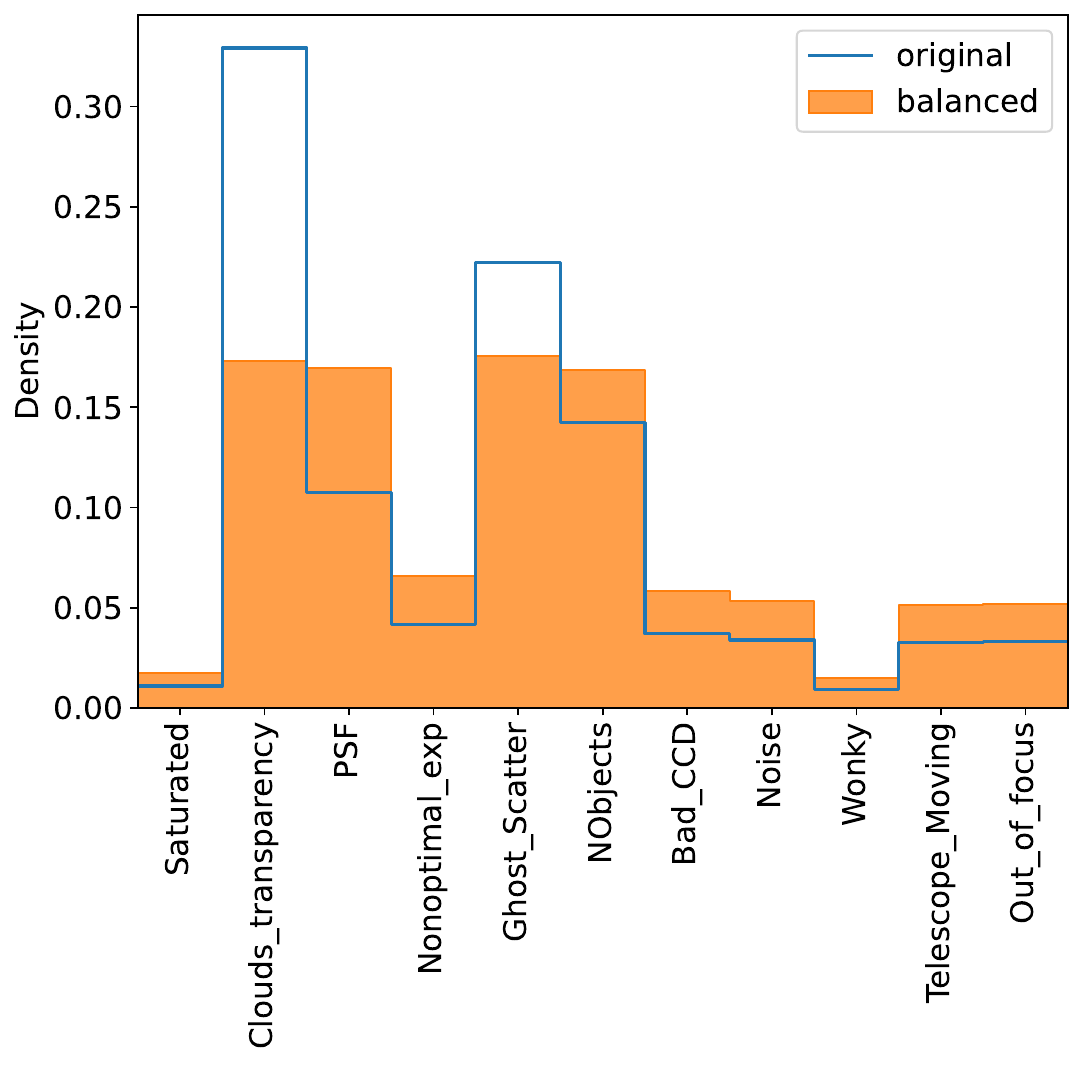}
    \caption{The comparison between the original sample of images and the balanced dataset in each category. The ``Density'' depicts the normalized count of images. }
    \label{fig:balance_comparison}
\end{figure}

The kNN training described in Section~\ref{subsec:knn} requires a relatively balanced number of images in each category for a good predictive performance.
The original labeled dataset is very imbalanced with the ratio of the number of images in the most populated (\textit{Clouds\_transparency}) and least populated (\textit{Wonky}) categories being 35.
To remedy the imbalance, we downsampled the three categories with the highest number of images and subsequently upscaled the weight of those three categories in the kNN training.
Specifically, we downsampled \textit{Clouds\_transparency} to one-third of its original size (and correspondingly upweighted the category by 3),
downsampled \textit{Ghost\_Scatter} by half, and
downsampled \textit{NObjects} by 0.75 (increasing the weight by 1.333).
With this downsampling and upweighting procedure, the ratio of imbalances between images reduces to 11.8, almost  3-times lower than the original ratio.
The comparison between the original and balanced datasets is shown in Figure \ref{fig:balance_comparison}.

\subsection{Summary}

Our dataset draws from DECaLS imaging in LS DR9 and DR10, supplemented with exposures from DELVE. Problematic exposures are identified from four main sources, which we summarize below (see also Table~\ref{tab:datasetCompose}):

\begin{enumerate}
    \item The DECaLS Bad Exposure List, which we use to flag poor quality exposures, bad observing conditions, or instrument issues.
    \item DES Expert Labels. These are visual inspections from the DES Y6 release.
    \item DELVE Expert Labels. These are visual inspections from DELVE DR3.
    \item LS pipeline flags. These are automated quality flags applied during LS pipeline processing, which we use to prune further potentially bad exposures from the good exposure candidate set.
\end{enumerate}

We unified the labels from the DECaLS and human expert lists into the 11 categories described in Sec.\ \ref{subsec:pattern}, focusing on the most common issues that can be generalized across exposures.
We reduced the dataset to these categories to improve model generalization, downsampled overrepresented classes, and balanced the good/bad exposure ratio for classification. The final labeled set contains 73{,}990 bad exposures and 10{,}000 good exposures.

\section{Method}\label{sec:method}

The semi-supervised approach means that our pipeline employs both supervised and unsupervised learning.
The two main components in our pipeline are a Deep Neural Network (DNN)  trained via Self-Supervised Learning (SSL) \citep{ssl_intro, gui2024surveyselfsupervisedlearningalgorithms} for pattern recognition and embedding generation, and a classifier for categorizing the images based on the generated embeddings, trained using supervised learning with labels.

%SSL emerges as a powerful paradigm to address the challenges of bad exposure identification.
%SSL is a powerful method for identifying bad exposures as it offers the potential to eventually eliminate the need for labeling by human experts, a process that is time-consuming, expensive, and increasingly impractical for large-scale astronomical datasets.

%It also has the potential for discovering of Nuanced Features.
%Beyond mere automation, 
SSL demonstrates remarkable potential for uncovering subtle, previously unknown features within data, as well as for removing human subjectivity from the labeling process. These qualities may be valuable when identifying bad exposures, as human experts often merge or confuse similar categories of problems (cf. Section \ref{subsec:expertLabel} where we discuss collapsing 61 different categories down to 10), or entirely miss important features.
SSL is also scalable, which should be useful as astronomical surveys continue to grow in size and complexity.
%As astronomical surveys continue to grow in scale and complexity, SSL algorithms can adapt and scale accordingly, maintaining efficiency and accuracy across diverse datasets.
%Regarding consistency, by removing human subjectivity from the labeling process, SSL can provide more consistent and reproducible results in bad exposure identification.
% Finally in regards to Continuous Learning, SSL models can potentially be designed to adapt and improve their performance as they process more data, staying current with evolving observing conditions and instrument characteristics.

Perhaps most importantly, SSL offers the potential for adaptation
%While SSL itself doesn't inherently provide continuous learning, it can serve as a foundation for developing adaptive systems.
because it facilitates easier retraining compared to supervised methods, as it doesn't require manual labeling of new data. 
%This can be particularly advantageous in astronomy, where the definition of a "bad exposure" might evolve over time due to changes in instrumentation, observing strategies, or scientific priorities.
%By periodically retraining SSL models on new data, astronomers can potentially keep pace with evolving observing conditions, instrument characteristics, and newly discovered types of bad exposures.
For example, a SSL model trained on current data could be periodically fine-tuned on more recent observations. This process could help the model adapt to subtle changes in image characteristics or new types of artifacts, all without requiring manual annotation of these new examples.

Given the benefits of SSL, our pipeline is built around a pretrained Vision Transformer (ViT) model \citep{ViT2020}, a powerful deep learning architecture that has shown impressive performance on a wide range of computer vision tasks \citep[e.g.,][]{sam_paper,vit_medical}.
Our pipeline comprises three sections: a Deep Neural Network -- ViT -- for pattern recognition and embedding generation, an embedding post-processor, and a classifier.
The input to the pipeline is a single CCD image and the output is a vector of probabilities associated with each issue category for the image.
The pipeline is tuned using Hyperparameter Optimization (HPO) to improve the prediction performance \citep[e.g.,][]{hpo_info}.

%Efficiency is a bottleneck in our implementation due to the excessive number of CCD images given each 
As each exposure comprises 61 or 62 CCD images as noted in Section \ref{sec:intro}, running each image through our pipeline is computationally demanding for both the training and inference stages of our classification process. Further, some features, such as transparency issues due to cloud, or unexpected movement of the entire telescope due to, e.g., wind shake, are large-scale and can affect multiple CCD images. This means that we have to design our model to balance efficiency and accuracy.
Therefore, for exposure-level information, we randomly draw 20 CCD images from each exposure and employ a ``voting consensus'' method. Basically, if 10 of the 20 CCDs are considered bad, then the entire exposure will be labeled as bad. We will discuss the details of the pipeline in the rest of this section.

\subsection{Pipeline Overview}
\label{subsec:knn}

Our pipeline starts by feeding an image from the image dataset into the Deep Neural Network to generate an embedding with 768 dimensions.
Some dimensions contain redundant or unstructured information (i.e.,\ noise). We use post-processing to mitigate this noise to improve the overall prediction performance.
The embedding post-processors include a scaler, a dimension reducer and a feature selector \citep{Maharana_2022}.
Output embeddings from the ViT are run through these post-processors before being passed to a kNN classifier.
For the final step, the kNN classifier takes the post-processed embeddings and assigns a label with a probability based on the distance to the nearest neighbors in the training dataset.
By comparing the probabilities and the labels, we can determine the qualities of the image from two perspectives: 1. {\em whether} the image is good or bad; 2. {\em if} the image is bad, which bad category it belongs to.
Our end-to-end pipeline is illustrated in Figure~\ref{fig:pipeline}.

\tikzstyle{startstop} = [rectangle, rounded corners, minimum width=1cm, minimum height=1cm,text centered, draw=black, top color=white, align=center]
\tikzstyle{process} = [rectangle, minimum width=1cm, minimum height=1cm, text centered, draw=black, top color=white, bottom color=blue!20, align=center]
\tikzstyle{decision} = [diamond, minimum width=1cm, minimum height=1cm, text centered, draw=black, top color=white, align=center]
\tikzstyle{arrow} = [thick,->,>=stealth]

% Define colors
\definecolor{lightblue}{RGB}{173, 216, 230}
\definecolor{lightgreen}{RGB}{144, 238, 144}
\definecolor{lightred}{RGB}{255, 182, 193}
\definecolor{lightgray}{RGB}{200, 200, 200}

\begin{figure}
\centering
\begin{tikzpicture}[node distance=2cm, auto]
    \node (dataset) [startstop, align=center] {Image\\ Dataset};
    \node (vit) [circle, draw, right of=dataset, bottom color=lightgray, align=center, xshift=0.5cm] {ViT\\ Model};
    \node (post) [process, right of=vit, align=center, xshift=1cm] {Post-\\processor};
    \node (clustering) [startstop, above of=post, align=center] {Clustering\\Analysis};
    \node (knn) [process, right of=post, align=center, xshift=1cm] {kNN\\ Classifier};
    \node (decision) [decision, right of=knn, align=center, bottom color=yellow, xshift=3cm] {label $\neq$ 0\\and\\Prob $>$ 0.9?};
    \node (bad) [process, below of=decision, bottom color=lightred, yshift=-1.5cm] {Bad};
    \node (good) [startstop, above of=decision, bottom color=lightgreen, yshift=1.5cm] {Good};
    \node (badcat) [startstop, xshift=-3cm, left of=bad, top color=lightred, bottom color=lightblue] {Bad Category};
    \draw [arrow] (dataset) -- (vit);
    \draw [arrow] (vit) -- node[above] {} (post);
    \draw [arrow] (post) -- node[above] {} (knn);
    \draw [arrow] (post) -- node[above] {} (clustering);
    \draw [arrow] (knn) -- node[above] {Probability} node[below] {label} (decision);
    \draw [arrow] (decision) -- node[left] {Yes} (bad);
    \draw [arrow] (decision) -- node[left] {No} (good);
    \draw [arrow] (bad) -- node[above] {Assign label} node[below] {with probability} (badcat);
\end{tikzpicture}
\caption{A depiction of how our pipeline identifies bad exposures.}
\label{fig:pipeline}
\end{figure}
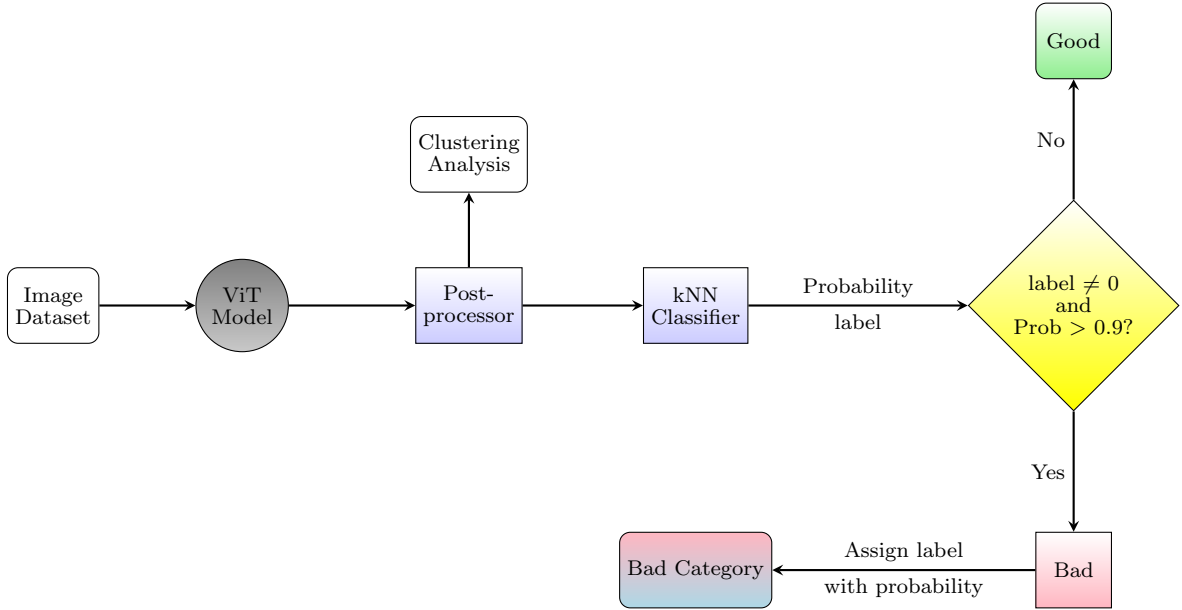

% \includegraphics[width=0.4\textwidth]{figures/pipeline.png}

% \subsection{The Vision Transformer}\label{subsec:vit}

%\yl{Introduction to DINO} The self-DIstillation with No Labels (DINO, \citealt{caron2021emerging}) is a self-supervised learning framework that trains a deep learning model on a large dataset for pattern recognition without human-annotated labels. In the framework, a "student" model, with a partial view of an image, learns to match the predictions of a "teacher" model that sees the full image. This sharpens the student’s pattern-recognition skills, while the teacher improves by incorporating the student’s progress, forming a continuous improvement loop. To avoid oversimplification and thus knowledge collapsing, DINO uses data augmentations, such as centering to prevent narrow focus and sharpening to preserve detail, ensuring balance and stability of the learning process.

%Unlike other methods, DINO doesn’t depend on large data batches, enhancing flexibility and scalability. By refining information processing, it excels at image recognition without labeled data, trading some flexibility for stability and efficiency, as proven in experiments. This student-teacher dynamic and stabilization create a robust framework for capturing visual representations.

% \adm{this diagram isn't currently referenced in the text. I think we should remove it.}
% \begin{figure}
% \centering
% \includegraphics[width=0.3\textwidth]{figures/DINO_module.png}
% \caption{The DINO workflow diagram. x is the original image. x1 and x2 are two augmented view of x.}
% \label{fig:DINO}
% \end{figure}

The Deep Neural Network we employ is a ViT \citep{ViT2020} trained on ImageNet \citep{ImageNet} using the DINOv2 framework \citep{oquab_dinov2_2023}.
Specifically, the ViT model we use is the \code{vit\_base} model with a register instance \citep{darcet2023vitneedreg}, downloaded from PyTorch Hub, the specific model is
\begin{verbatim}
"facebookresearch/dinov2","dinov2_vitb14_reg"
\end{verbatim}
The \code{vit\_base} model is an 18-block Vision Transformer with 768-dimensional embeddings and 12 attention heads.
Similar to a conventional transformer model for Natural Language Processing (NLP), the ViT model is composed of a stack of transformer blocks.
Each input image is divided into a grid of patches --- 168x84 square patches each with 14x14 pixels in our chosen model --- and then transformed into a sequence of vectors.
Paddings with zeros are used to ensure the images are divisible by the patch size.
This sequence of vectors is then fed into the transformer blocks to generate the embeddings. See \citet{darcet2023vitneedreg}, and references therein, for more details.
% An workflow diagram is shown in figure 
% \begin{figure}
%     \centering
%     \includegraphics[width=0.4\textwidth]{figures/vit.pdf}
% \end{figure}

% \subsection{k-Nearest Neighbor Classifier}\label{subsec:knn}

The k-Nearest Neighbor (kNN) classifier is a simple and effective algorithm for classification tasks based on the distance between the input data and the training data.
We employ the \code{scikit-learn} implementation \citep{scikit-learn} of the kNN classifier for its simplicity and efficiency.
The labeled data provides the training set for the kNN classifier.

\subsection{Training and Hyperparameter Optimization}\label{subsec:training}

Since we used a pretrained ViT model for the embedding generation, our focus in this paper is on training the downstream pipeline --- the embedding post-processor and the kNN classifier.
After the ViT model generates the embeddings for both the training and test datasets, we used the embeddings to {\em simultaneously} train the post-processor and the kNN classifier.

We employed the HPO technique \citep[e.g.,][]{hpo_info} to search for the optimal hyperparameters for the post-processor and the kNN classifier.
In particular, we defined ranges (for continuous variables) and options (for discrete variables) for hyperparameters and used the {\em Random Search with Sucessive Halving (RSSH)} algorithm \citep{sh1,sh2} to search for the optimal hyperparameters.
The ranges and options are chosen to be broad to let the \textit{RSSH} algorithm explore for more globally optimal parameters.
We also used 3-fold CV to evaluate the performance of the pipeline to ensure that our pipeline can be well generalized to a new dataset.
Since our task is a multi-class classification problem, we used the overall accuracy (see Equation \ref{eqn:acc}) as the metric to evaluate the performance of the pipeline during HPO -- the higher the accuracy, the better the hyperparameter combination.
The hyperparameter search space is summarized in Table \ref{tab:hpo_space}.

\begin{table}[h]
    \centering
    \begin{tabular}{c c}
        \hline
        Component & Hyperparameter Search Space \\
        \hline \\
        \multirow{2}{*}{Scaler} & StandardScaler, \textbf{MinMaxScaler}, Normalizer, \\
        & MaxAbsScaler, RobustScaler, PowerTransformer \\
        PCA number of components & 5, 10, \textbf{15}, 25, 30, 50, 70, 100 \\
	    VarianceThreshold & 0, \textbf{0.001}, 0.01 \\
	    kNN number of neighbors & 1, 3, 5, \textbf{7}, 10, 20, 30 \\
	    kNN distance metric (p) & \textbf{1}, 2, 3, 5 \\
        kNN leaf size & \textbf{1}, 5, 10, 15, 30, 35 \\
        \hline
    \end{tabular}
    \caption{The hyperparameter search space for HPO. The bold values indicate the determined optimal hyperparameters.}
    % with highest multiclass accuracy.}
    \label{tab:hpo_space}
\end{table}

The core code components employed by our technique, and their associated optimal hyperparameters, are as follows.
We used the \code{MinMaxScaler} to fix the range of each dimension in the embedding vector to [0, 1].
We used Principle Component Analysis (PCA) with 15 components to reduce embeddings' dimensions from 384 to 15.
We then further chose \code{VarianceThreshold} as our feature selector to remove insignificant (or ``noisy'') features in the dimension-reduced embeddings with threshold=0.001.
Finally, we found that the kNN classifier performed optimally using 7 neighbors and the ``Manhattan distance'' as the distance metric.
The Manhattan distance\footnote{\url{https://mathworld.wolfram.com/TaxicabMetric.html}} (or L1 distance) is defined as:
$$d_{L1} = \sum_{i=0}^{dim} \left|p_i - q_i\right|$$
where $p_i$ and $q_i$ are the $i$-th dimension of the two embedding vectors $p$ and $q$ summed over $dim$ dimensions (i.e.\ essentially the sum of the absolute differences across the vector components). ``Using 7 neighbors'' in this sense means that the label and probability of each {\em test} image is determined by the distance to the 7 nearest neighbors in the {\em training} set.

\subsection{Inference}\label{subsec:inference}

After training and evaluation, we applied our pipeline to exposures that appear in DR11 of the LS (see Section \ref{sec:dataset}) to infer which exposures might belong to the categories listed in Table~\ref{tab:category}. DR11 contains 283{,}633 exposures, so the inference step is somewhat computationally taxing.  We therefore split the DR11 exposures into multiple batches and conducted the inference step in parallel on the National Energy Research Scientific Computing (NERSC) Center's Perlmutter\footnote{\url{https://docs.nersc.gov/systems/perlmutter/architecture}} supercomputer using 2 nodes with 4 GPUs each. Fortunately, since the inference step doesn't have to share gradients, it can be run on completely independent GPUs. Therefore, we can rapidly process many images in parallel with little-to-no performance degradation.

\section{Results}\label{sec:results}

In this section, we present the performance of our trained pipeline on the DR10 data and inference results for the DR11 candidate exposures.
We evaluate our results in two ways.
First, a clustering analysis is performed to ensure images with similar features can be recognized by the model and separated in a cluster.
We specifically select the \textit{good} vs.\ \textit{Ghost\_Scatter} categories as a case study to illustrate how clusters are selected and interpreted by our pipeline.

Then, we evaluate our classification results using the multiclass \textit{accuracy}, \textit{precision}, \textit{recall} and a more balanced accuracy referred to as the \textit{F1 score} \citep{F1_intro}.
In the equations below, the TP, TN, FP, and FN are the number of true positive, true negative, false positive, and false negative classifications, respectively.
The multiclass accuracy is defined as \citep{grandini2020metricsmulticlassclassificationoverview}:
\begin{equation}
    \text{Accuracy}_i = \frac{\text{TP}_i}{\text{the number of images with true label}_i} \label{eqn:acc}
\end{equation}
The precision, recall and F1 scores are defined with respect to each class individually, so we follow the definition:
\begin{eqnarray}
    \text{Precision} = \frac{\text{TP}}{\text{TP} + \text{FP}} \label{eqn:precision} \\
    \text{Recall} = \frac{\text{TP}}{\text{TP} + \text{FN}} \label{eqn:recall}
\end{eqnarray}
where, the F1 score is then defined as:
\begin{eqnarray}
    \text{F1} = 2 * \frac{\text{Precision} * \text{Recall}}{\text{Precision} + \text{Recall}} \label{eqn:f1}
\end{eqnarray}

We use the accuracy to measure the overall performance of the classifier, while we employ the F1 score to measure the balance between the {\em precision} and the {\em recall}.
%; in other words, it can be treated as a balanced accuracy.

The good exposures significantly outnumber the bad exposures, so the model has to capture imbalanced numbers across different categories.
In this case, the raw {\em precision} of the model can better reflect the classification performance, as {\em precision} focuses on the classifications for the positive results (i.e.,\ the categories to which the model actually assigns labels).

Beyond the results we will report in this section, it is worth noting that our pipeline is also very fast due to our parallelization across GPUs.
For training, the ViT is already pre-trained and, our focus is on training the post-processor and the kNN classifier.
The HPO process took a total of 2.5 hours to complete for our predefined set of parameters on a single node with a 128-core CPU.
The only bottleneck in our pipeline is during the inference stage, when generating the embeddings using the ViT. Once the embeddings are generated, the kNN utilizes them without further intensive processing. To better optimize the embedding generation task, we distributed the inference tasks to 4 nodes, each with 4 GPUs.
Overall, the inference step applied to all 5{,}672{,}660 LS DR11 images (283{,}633 exposures with 20 CCD images drawn from each exposure, as described in Section \ref{sec:method}) took a total of only 41 hours. On average, each GPU can process and generate embeddings for 8{,}647 images per hour.

\subsection{Clustering Analysis} \label{sec:clustering}

For a SSL algorithm, the performance is usually evaluated first by qualitatively checking how the embeddings are clustered to see if there are clear separations and structures among different potential characteristics of the images.
%Each cluster usually represents a different characteristic of the images.
A clear separation indicates that the model has definitively learned certain features of the images and is able to distinguish distinct patterns.
This will ensure that the downstream task --- the classification of new bad exposures --- will have a good performance baseline (see Section \ref{subsubsec:case_study} for more details).

\begin{figure}
    \centering
    \includegraphics[width=0.9\textwidth]{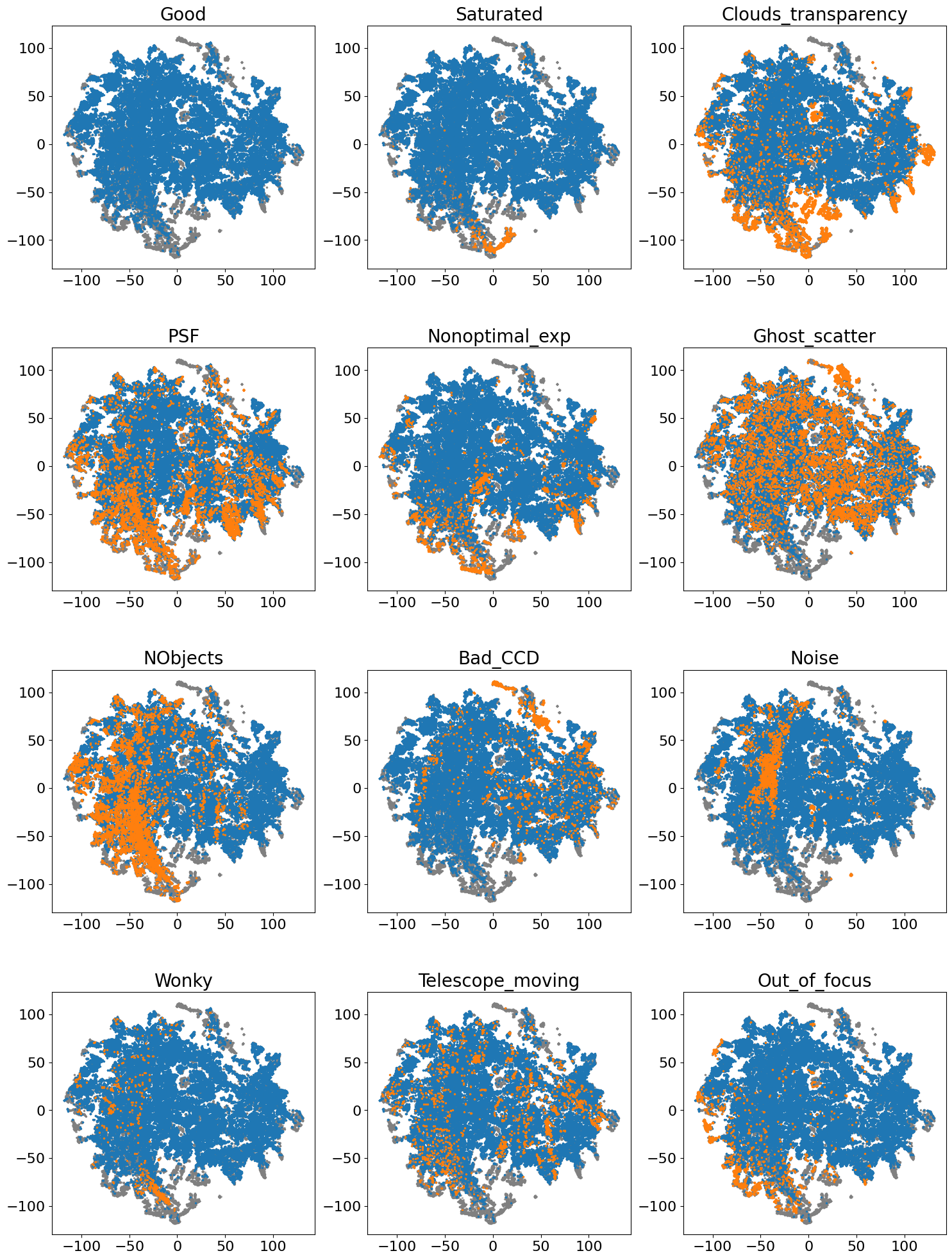}
    \caption{The clustering of the embeddings generated by the ViT model for the training dataset. The embeddings are processed through the same data processors described in Section \ref{subsec:training}, and they are further dimensionally reduced using the t-SNE method \citep{tsne} to help with visualization. The grey dots represent all training datasets, the blue dots depict ``good'' exposures, and the orange dots highlight the labeled exposures for each ``bad'' category. The top-left panel displays the good exposures, and the other panels show the bad exposures. The axes of the figures correspond to two dimensions of the t-SNE-reduced embeddings, and are not physically interpretable.}
    \label{fig:clustering}
\end{figure}

%The purpose of the clustering analysis is to understand how well the model has learned the features of the images and the capability of distinguishing the images using these features.
We focus on using the training dataset to evaluate the clustering results, as it has known labels. The separation of the clusters is most easily visualized after running the embeddings through the dimension reduction process, the results of which are shown in Figure \ref{fig:clustering}.
Generally, well-separated and tightly-bound clusters indicate the model has learned enough features to distinguish different categories.
We want to emphasize that plots such as Figure~\ref{fig:clustering} -- t-Distributed Stochastic Neighbor Embedding~\citep[t-SNE;][]{tsne} plots are only useful for qualitative analysis, and that the displayed clustering results have been collapsed over many dimensions. This means that the axes of the figures, and the coordinates of each single data point, are not intuitively {\em quantitatively} interpretable in and of themselves.
%The data points must be analyzed in a cluster with other data together.

Figure \ref{fig:clustering} demonstrates that there is a clear clustering structure for the training dataset.
As a more quantitative metric, the ``Silhouette Score''\footnote{See \url{https://scikit-learn.org/stable/modules/generated/sklearn.metrics.silhouette_score.html}} of the 
clustering is 0.23, showing some indications of clustering.
There are several distinct areas that represent different categories of exposures.
Most of the good exposures occupy the central parts of the space, while the bad exposures separate out across many different clusters.
The good exposures typically contain various noticeable pattens that the model may pay attention to in a manner that a human visual inspector would not.
Most prominently, extinction caused by Galactic dust is easily identified by the model as a distinct feature. The model therefore incorrectly identifies dusty exposures as ``bad''.
We alleviate this issue for our final results by applying an extinction cut to the ``good'' exposure candidates to avoid dusty regions as described in Section \ref{subsec:extCut}.

The wide spread of the ``good'' exposures indicates that the corresponding
features are very diverse and span an extensive range of the embedding space,
as one would expect from a large imaging survey.
Due to the high complexity of the images, the weak separation of clusters is mainly caused by the large scattering of ``good'' exposures.
We will discuss how a probability cut in Sec. \ref{subsec:classification} can reduce the effect of this scattering. 
For the ``bad'' exposures, the ``ghost'' exposures and the exposures in the ``NObjects'' category are 
spread out and overlap with the clusters of ``good'' exposures.
It is worth remembering that Figure \ref{fig:clustering} displays a two dimensional reduction of the embeddings, so overlapping clusters might actually be quite distinct in higher dimensions.
We compare the ``good'' and ``ghost'' exposures in the next sub-section to better characterize the separation of the clusters in higher 
dimensions and to illustrate the capability of the model to distinguish different features of the images.

For the other labels, the ``bad'' exposures are concentrated around certain parts of Figure \ref{fig:clustering} (i.e.,\ they have low spread) and some of the categories separate well from the ``good'' exposures.
A clear separation and a tight cluster indicate that the model has learned the features of the images well and should be able to distinguish patterns that correspond to different categories. These distinct patterns also provide a good basis for classifying test exposures as ``good'' or ``bad,'' as discussed in Section \ref{subsec:classification}.

\subsubsection{Case Study: {\tt Ghost\_Scatter} Exposures}\label{subsubsec:case_study}

\begin{figure}
    \centering
    \includegraphics[width=1.0\textwidth]{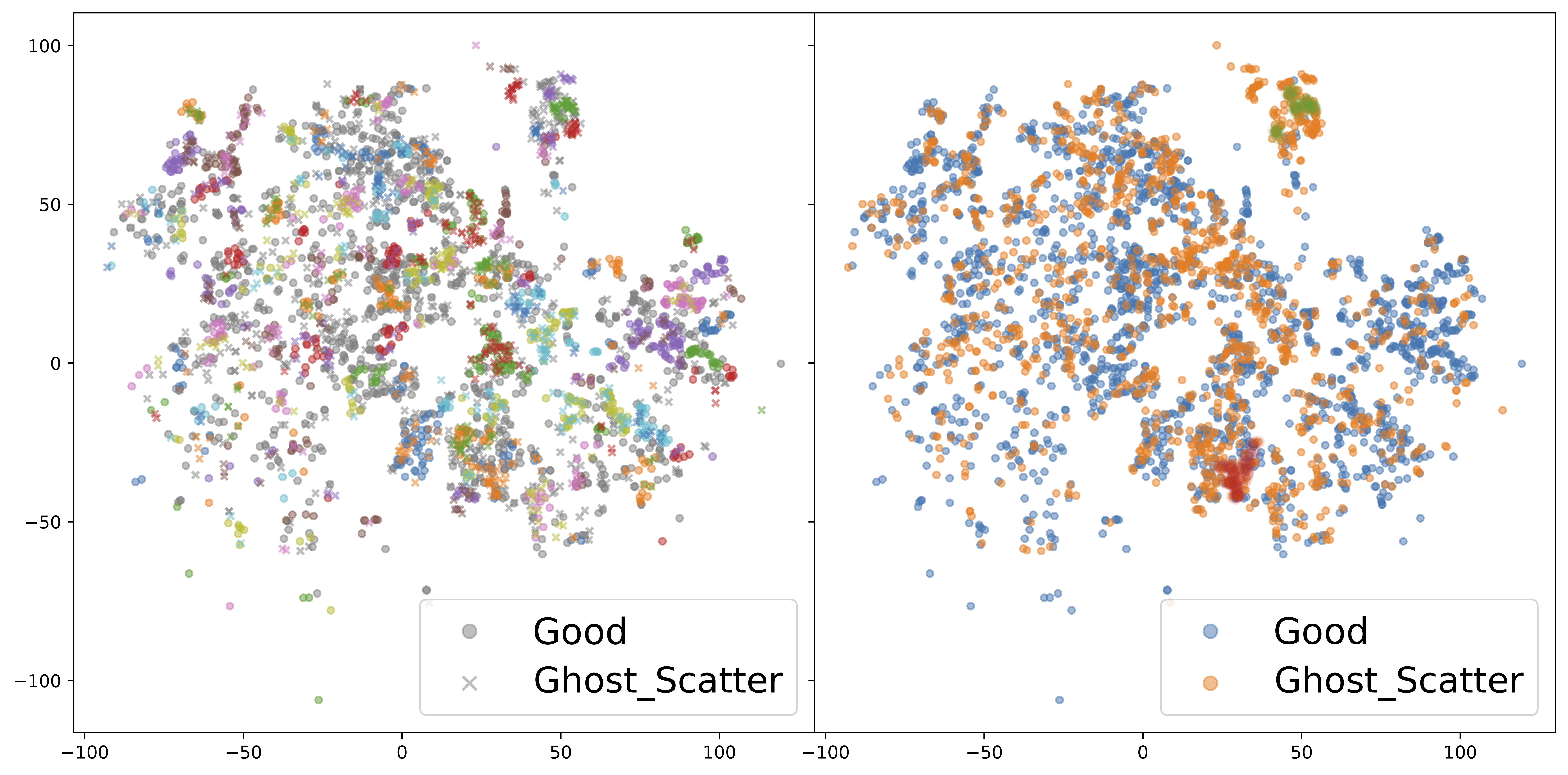}
    \caption{The left-hand panel depicts all clusters identified by the HDBSCAN clustering algorithm. The gray dots in the panel do not belong to any cluster. The right-hand panel highlights exposures in the \textit{Ghost\_Scatter} category and how they compare to ``good'' exposures. The red and green points displayed in this panel correspond to exposures displayed in Figure~\ref{fig:sample-exp}.}
    \label{fig:hdbscan}
\end{figure}

\begin{figure}
    \centering
    \href{https://brookluo.github.io/projects/bad_exposure/}
    {\includegraphics[width=0.9\linewidth]{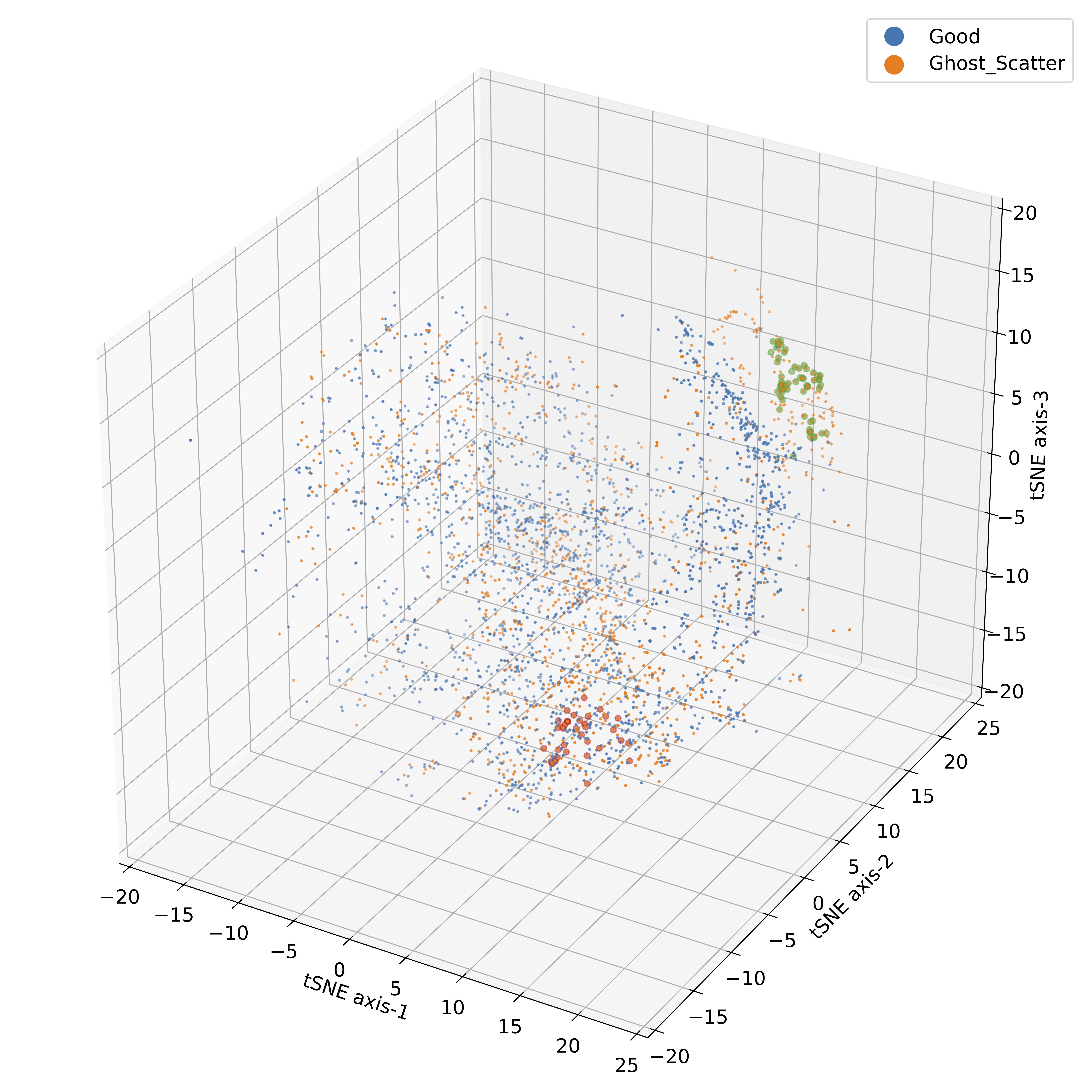}}

    \caption{Three-dimensional t-SNE plot comparing exposures in the ``good'' and \textit{Ghost\_Scatter} categories with higher dimensionality than for Figure \ref{fig:clustering}. The red and green points depicted in this figure correspond to exposures displayed in Figure~\ref{fig:sample-exp}. This plot can be viewed interactively at \url{https://brookluo.github.io/projects/bad_exposure/}, or by clicking on the figure.}
    \label{fig:tsne-3D}
\end{figure}

We applied the Hierarchical Density-Based Spatial Clustering of Applications with Noise~\citep[HDBSCAN;][]{hdbscan_theory,hdbscan_impl} algorithm to a randomly-drawn sample of 10{,}000 CCD images from our whole dataset to study how our derived embeddings cluster (see Figure \ref{fig:hdbscan}). We chose 10{,}000 images to reduce the computational cost for such a clustering analysis while keeping the sample size large enough to be statistically representative.
HDBSCAN identified 70 clusters that included more than 10 data points. The distinctness of these clusters is a good indicator of how well individual exposures will be able to be separated into different categories. We also compare exposures via a 3D t-SNE plot in Figure \ref{fig:tsne-3D}, which provides additional information about the spatial distance and separations of clusters.

We focus on the \textit{Ghost\_Scatter} category as a case study to demonstrate what Figures \ref{fig:hdbscan} and \ref{fig:tsne-3D} illustrate. The fact that the embeddings that correspond to the \textit{Ghost\_Scatter} category are almost entirely clustered in the top-right corners of Figures~\ref{fig:hdbscan} and \ref{fig:tsne-3D} demonstrates that the model has identified some unique features for these categories of exposures.
HDBSCAN identified several clusters in this region, however, which is evidence that other CCD images share properties with exposures in the \textit{Ghost\_Scatter} category. To try to ascertain {\em which} types of images share properties with exposures in the \textit{Ghost\_Scatter} category, we selected two clusters based on the HDBSCAN clustering results for further analysis. The clusters are depicted in green and red in Figure \ref{fig:tsne-3D}. We randomly draw three example exposures from each of the two clusters, and present them in Figure \ref{fig:sample-exp}. In this figure, we can observe that pupil ghosts exist in all of these exposures except for panel (a). A qualitative interpretation of the differences in the exposures is that the red \textit{Ghost\_Scatter} cluster from Figure \ref{fig:tsne-3D} is more diffuse, and mixes with the good exposures, so the CCD images in exposures drawn from this cluster might be ``slightly better'' (more similar to a good exposure). On the other hand, exposures that lie in the green cluster in Figure \ref{fig:tsne-3D} all show very clear, or even multiple, pupil ghosts.

%The difference of these two clusters can be interpreted in two ways.
%First, exposures with ghosting issue can be identified using our pipeline.
%Conversely, it also exposes that the current pipeline might suffer from issues where the pupil ghosts is small and localized.
%We will have a detailed discussion on this issue in Section \ref{subsec:future}.

% YL: 

\subsection{Classification Performance}\label{subsec:classification}

\begin{table}[h]
    \centering
    \begin{tabular}{|c c c c c|}
        \hline
        Category & Precision & Recall & F1-score & Support \\
        \hline
        \hline
        \textit{Good} & 0.970 & 0.764 & 0.854 & 2411 \\
        \hline
        \textit{Saturated} & 0.990 & 1.000 & 0.990 & 411 \\
        \hline
        \textit{Cloud\_transparency} & 0.977 & 0.990 & 0.981 & 3076 \\
        \hline
        \textit{PSF} & 0.872 & 0.960 & 0.915 & 940 \\
        \hline
        \textit{Nonoptimal\_exp} & 0.925 & 0.970 & 0.947 & 599 \\
        \hline
        \textit{Ghost\_scatter} & 0.833 & 0.910 & 0.868 & 693 \\
        \hline
        \textit{NObjects} & 0.824 & 0.960 & 0.886 & 283 \\
        \hline
        \textit{Bad\_CCD} & 0.949 & 0.980 & 0.963 & 1149 \\
        \hline
        \textit{Noise} & 0.842 & 1.000 & 0.913 & 461\\
        \hline
        \textit{Wonky} & 0.814 & 1.000 & 0.899 & 49 \\
        \hline
        \textit{Telescope\_moving} & 0.804 & 0.950 & 0.870 & 284 \\
        \hline
        \textit{Out\_of\_focus} & 0.974 & 0.990 & 0.981 & 874 \\
        \hline
    \end{tabular}
    \caption{The classification performance evaluated using three scores. The rightmost column (``Support'') records how many test images in a given category were used in the classification evaluation. This table is generated from the \code{classification\_report} function in the \code{scikit-learn} package.
    We re-ran our analysis with 10 sets of random seeds and found that the standard deviation of each metric is less than 0.025, indicating the stability of our model.
    }
    \label{tab:classification}
\end{table}

The general separation between clusters noted in Section~\ref{sec:clustering} suggests that it is reasonable to proceed to classify exposures using our model.
We therefore train and evaluate the kNN classifier discussed in Section~\ref{subsec:knn} by applying it to the post-processed embeddings from samples described in Section \ref{sec:dataset}.
The classifier can predict the label with a probability associated with the distance to the nearest neighbors in the training dataset.
We calculate this distance using the Manhattan distance metric, as described in Section \ref{subsec:training}.
The number of nearest neighbors in this case is a hyperparameter, which is chosen to be 7 as it is the optimal parameter determined by the HPO process outlined in Section \ref{subsec:training}.
For the multiclass classification, the probability of an image's label in each category is calculated by the number of neighbors in the same category divided by the total number of neighbors:
$$P_{j=0...11} = \frac{N_{{\rm neigh}=j}}{N_{\rm tot}} = \frac{N_{{\rm neigh}=j}}{7}$$
The numeric class labels are from 0 to 11 with 0 representing a good exposure and other labels representing each bad category.

\begin{figure}
    \centering
    \includegraphics[width=0.8\textwidth]{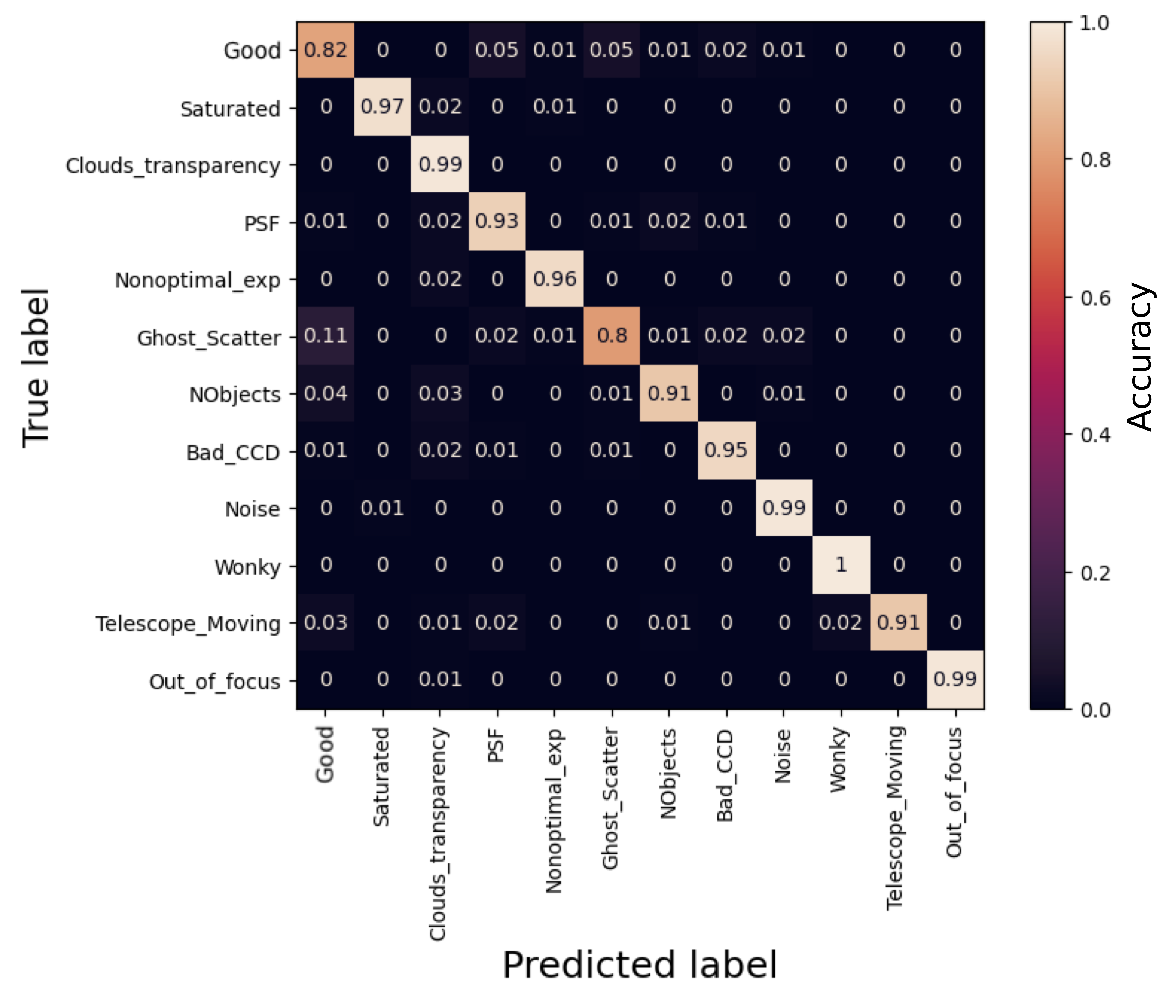}
    \caption{The confusion matrix for the multi-class classification problem. The displayed fractions correspond to the {\em multiclass accuracy} (see Equation~\ref{eqn:acc}). %We applied a probability threshold of 0.9 for the kNN classifier to improve precision. The fraction in each tile is the proportion of images predicted in that tile in the whole true label category. The tiles on the diagonal are the correct predictions made by the model.
    }
    \label{fig:confusion-matrix}
\end{figure}

After calculating the probability for each test exposure, we applied a probability cut of $\geq 0.9$ to improve prediction accuracy for our classification results.
Our final classification results are evaluated using accuracy, precision, recall and F1-score. 
The accuracy information is shown in Figure \ref{fig:confusion-matrix} as a confusion matrix.
The precision, recall, and F1-score are shown in Table \ref{tab:classification}.

From the confusion matrix, our model achieves a high multiclass accuracy (see Equation~\ref{eqn:acc}) of higher than 0.8 for 9 out of our 11 categories of bad exposures.
Table \ref{tab:classification} shows that our model has a consistent classification performance across 11 categories.
It is worth pointing out that \textit{Wonky} has a limited number of images in the testing dataset, which can be non-representative, and thus will affect the classification result for this category.
However, we expect exposures corresponding to the \textit{Wonky} category to be generally rare in any case, as we usually know the precise category that a bad exposure belongs to.

The good performance demonstrated by our model across multiple evaluation metrics is due to several aspects of our approach.
Fundamentally, we have confirmed in Section \ref{sec:clustering} that our model is able to recognize and distinguish different patterns.
Furthermore, the 11 bad categories we use as labels are sufficiently representative to cover a wide range of photometric and technical issues.
This may also imply that our model should be generalizable to other telescopes.
Lastly, we applied a cut of $>0.9$ in probability for our classification metrics, meaning we only focus on the most problematic exposures, which will inevitably improve the accuracy of our technique. 

The two categories for which our model performed worst are the \textit{Ghost\_Scatter} and \textit{Telescope\_Moving} categories.
The poor performance for \textit{Ghost\_Scatter} is likely due to ghost images spanning multiple CCDs, which we discuss more in Section~\ref{subsec:future}. This is in contrast to most ``bad'' categories, which impact an entire exposure and so produce a pattern that is noticeable in every CCD image (regardless of which set of CCDs is randomly chosen by our census voting method discussed in Section~\ref{sec:method}).
The low accuracy for the \textit{Telescope\_Moving} category is likely caused by this category having a relatively small training sample from which the classifier could learn.

\subsection{Inference and Visual Inspection} \label{sec:inferenceVI}

We ran our trained model on new exposures from the DECaLS and DELVE to determine which of these new exposures are consistent with being in a ``bad'' category, 
%We perform  are performed with the trained network and using four GPUs on Perlmutter cluster at NERSC. 
as detailed in Section \ref{subsec:inference}.
As derived in Section \ref{subsec:classification}, we applied a cut of $\geq 0.9$ in probability to determine a ``bad'' exposure. This resulted in a sample of 4436 high-probability bad-exposure candidates.

Our ultimate goal is to identify new exposures to add to the {\em DECam Bad Exposure List} described in Section~\ref{subsec:badExpList}, with a view to augmenting that list for LS DR11.
So, after running our identification pipeline, we visually inspected the bad exposure candidates with a view to only excluding extremely bad exposures. There are two reasons for this choice.
First, because DECam has a wide field with 62 individual CCDs (or 61 CCDs as stated in Sec \ref{sec:intro}), many non-critical local issues can be identified by our pipeline as meaningful patterns.
Second, because reaching a uniform depth is a major goal of the LS, it is preferable to err on the side of keeping marginal exposures. 

Our visual inspection of the bad exposures was conducted by three expert members of the LS collaboration. Each expert evaluated two aspects of each exposure --- whether the exposure was indeed a bad exposure, and, if so, whether the category label assigned by the model was appropriate.
The human experts' responses were collated to form a final sample of 946 bad exposures. Some of these are existing exposures from LS DR9 and DR10 and some are identified only in new imaging assembled for DR11. All of the bad exposures will be rejected to produce the final reduced data for LS DR11.

\section{Discussion}\label{sec:discussion}

In this section, we further analyze the results from our bad exposure identification pipeline. To do so, we consider two different sets of exposures: those used for the final LS DR9 and DR10 datasets, and new exposures identified from the DECaLS and DELVE campaigns intended to be used for LS DR11 (cf.\ Section \ref{sec:dataset}).
We will discuss how some exposures in the LS DR10 dataset are identified as ``bad'' by our pipeline but were entirely missed by the LS pipeline (henceforth {\tt legacypipe}).
%We will discuss these exposures and compare them with the Legacy pipeline's results.
%For the new exposures, the highlights are put on the categories that the model considered it to be in.
Finally, we will also discuss the limitations of our current pipeline and some approaches that we might use to improve our current model and to adapt it to other imaging datasets.

\subsection{Comparison between our pipeline and {\tt legacypipe} for LS DR9 and DR10}

The {\tt legacypipe} is applied to data that has already been processed by the Community Pipeline \citep[CP;][see also Section~\ref{subsec:cpExtCut}]{CP_pub}. The CP performs a number of initial data-reduction steps, including bias and dark-current subtraction, flat-fielding and an initial background estimation. The {\tt legacypipe} then processes each image to obtain photometry by performing full source and background modeling. A problematic exposure could be flagged by either the CP or by {\tt legacypipe}. In addition, data reductions for DR9 and DR10 utilized a human-derived bad exposure list similar to that described in Section~\ref{subsec:badExpList}. However, some exposures we identified as outlined in Section~\ref{sec:inferenceVI} were not previously noticed by any of these LS reduction stages, and were included in LS DR9 or DR10. Our pipeline identified 2{,}591 candidates. Of these, only 579 exposures were also identified as bad by \lsp. During our subsequent visual inspection, human experts verified 67 bad exposures that were included in DR9, and 92 bad exposures that were included in DR10. Our model therefore complements previous methods used by the LS to reject bad exposures.

Figure \ref{fig:bad-exp-ghost} shows an example of a bad exposure included in LS DR9 that would have been rejected by our pipeline.
Our model classified this exposure as bad with label \textit{PSF}, which typically means that an exposure includes stars with an irregular PSF.
In actuality, this image does not have any noticeable PSF issues, but does contain prominent pupil ghosts. We expect that this incorrect label is due to the model's limited focus on using single CCD images to capture each label (see the discussion in Section \ref{subsec:future}), and that ghosting across a single CCD could resemble a bad PSF.
%The pixel-level information for this exposure suggests that the very bright sections of the images are prominent in several adjacent pixels within a CCD, which could resemble a bad PSF from pixel level.

\subsection{New Bad Exposures}

LS DR11 is expected to contain 107{,}940 new DECam exposures from the observing period 2020A -- 2024A. We applied our pipeline to these exposures after they had been processed by the CP, and identified 1{,}732 bad exposure candidates, which is about 1.6\% of the new exposures.
The subsequent visual inspection confirmed 780 bad exposures.
Running these exposures through the {\tt legacypipe}, we find 708 mutually identified bad exposures and 72 exposures identified by our pipeline that
are not flagged by {\tt legacypipe}.
We investigated these 72 exposures and discovered that they were mostly in the \gs, \textit{NObjects}, and \textit{Cloud\_Transparency} categories.

The exposures identified by our pipeline will be included in the {\em DECam Bad Exposure List} (see Section~\ref{subsec:badExpList}) that the {\tt legacypipe} will use to reject exposures from LS DR11.

%This discovery poses a valuable insight to improve and complement the current pipeline.
%The ghost pupil and scattered light are common issues around bright sources, such as stars.
%Although these features are unavoidable in many exposures, we should be able to remove CCDs affected severely. 

\subsection{The Benefits of DINO and SSL}

%The importance of detecting bad exposures in astronomical imaging campaigns is paramount because the size of the datasets necessary for conducting statistical analysis on imaging surveys turns the intervention of humans for data managing impractical.
%The impact of bad exposures on astronomical research can be significant, potentially introducing biases in statistical analyses and hindering the detection of rare or transient phenomena.

% Self supervised learning is a machine learning paradigm that dispenses of humans for labelling the data to be used for training.
% But self supervised learning utilization in this research goes beyond the fact of dispensing of human in the loop because this machine learning paradigm offers great potential for discovering nuanced features hidden in the data completely autonomously.

Our approach to bad exposure identification takes advantage of recent advances in SSL by utilizing the pretrained DINO (v2) framework \citep{caron2021emerging, oquab_dinov2_2023}. Rather than training a self-supervised model from {\em scratch} on astronomical imaging alone, we extract rich embedding vectors from DINO to represent our telescope images. A major advantage of this strategy is that the
DINO model, pretrained on ImageNet, has been exposed to a very diverse set of natural images, which allows the model to develop a more sophisticated understanding of image structure than would be possible through studying astronomical images alone. Leveraging a model that has already distilled information from millions of diverse images is also a computationally efficient way to add extra richness to our classification pipeline.

%This diversity paradoxically provides an advantage over training directly on telescope images.
%While astronomical images might appear simpler and more uniform, the rich feature representations learned from ImageNet's variability enable the model to capture subtle variations and anomalies in telescope images.
%The complexity and diversity of ImageNet training allows the model to develop a more sophisticated understanding of image structure, which can be transferred to our specific domain.

%We also have the benefits of pretext task design. 
The self-supervised pretext tasks used in DINO's training process also offer unique advantages over traditional supervised approaches \citep{albelwi_survey_2022,batutin_self-supervised_2024}.
While supervised learning methods are constrained by rigid label definitions, self-supervised pretext tasks (such as image augmentation recognition) enable the model to learn nuanced features that extend beyond simple classification boundaries and that otherwise might be overlooked. %in a supervised setting.

Overall, in this work, we have found that using a framework like DINO in the context of SSL is practical, efficient, and returns excellent results when trying to classify bad exposures from large imaging surveys.

\subsection{Limitations and Future Improvements} \label{subsec:future}

Although our current implementation can successfully identify bad exposures that are missed by other approaches, there are several ways in which our pipeline might be improved.
%First, some patterns are large-scale spanning across multiple CCDs. Therefore, spatial information about the pattern can be included to increase accuracy.
%If a model can process the adjacent CCDs simultaneously, issues such as ghost, fringing, and wonky will be more obvious for the model.
%This can be achieved by including additional positional encoding for ViT.

One potential failing of our methodology is that we treated the ``good'' exposures as merely being the complement of the ``bad'' exposures. Our rationale is that the good exposures do not have a generic set of patterns that can be used to identify them, and are thus difficult to describe, whereas the bad exposures have distinct features that can be categorized. However, a better approach might be to specifically classify both the bad and good exposures first, and then to further classify the bad exposures into subcategories. This could be achieved via a two-step classification process, with each step having an individual classifier.

Perhaps the critical drawback of our approach (cf.\ Section \ref{subsec:pattern}) is that our current technique focuses on information in a single CCD image and uses a census voting method based on drawing 20 randomly-chosen CCD images, to ascertain effects across a whole exposure. This is the natural approach to adopt because we individually process each CCD image. But, certain categories of bad exposures, such as the {\tt Ghost\_Scatter} category (see, e.g.,\ Section \ref{subsubsec:case_study}), include patterns that span multiple CCDs. %So, even though 20 images comprise almost one-third of an entire exposure, 
So, it is certainly possible that we misclassify some images as ``good'' that would clearly be considered ``bad'' based on coherent large-scale information.
One concrete example is shown in Figure \ref{fig:mis-tm}, which shows an exposure that our pipeline classified as bad in the \tm~category. However, this was subsequently flagged as a misclassification during visual inspection by human experts.
In the exposure, there are two discontinuous, parallel satellite trails. This likely confused our model as it is a novel feature; most satellite trails in our exposures are continuous across an entire exposure.
This, and similar, issues would likely be solved by a model that can analyze patterns that span multiple CCDs.

One way to better identify certain categories of bad exposure might therefore be to process the entire exposure at once. This would require more memory but, on the other hand, our current method is computationally expensive in processing each individual CCD. Another approach would be to process adjacent CCDs in bunches, by including additional positional encoding in the ViT. Perhaps ideally for our purposes, there are ML-based methods of {\em hierarchical} image learning for high-resolution images, such as the Hierarchical Image Pyramid Transformer \citep[HIPT;][]{chen2022scalingvisiontransformersgigapixel}. HIPT potentially has the capability to process a whole exposure to discover large-scale patterns while simultaneously utilizing positional information to extract patterns on scales that correspond to one, or a few, CCDs.

Another issue is the current pipeline still requires human intervention to visually inspect the bad exposures identified by our model. Although this step is time-consuming, it is necessary to ensure that the bad exposures are indeed problematic, and to verify that the category assigned by our model is appropriate.
This issue could be addressed if a general consensus could be reached on the definition of a bad exposure.
With a larger set of bad exposures and a concrete standard for bad exposures, it might be possible to train a more sophisticated model that can classify bad exposures without any human intervention.

%Many new deep learning models, like Hierarchical Image Pyramid Transformer (HIPT), can address some of the issues above.
%These new models can be promising given the data volume and processing power are increasing exponentially.

\section{Conclusion}\label{sec:conclusion}

In this paper, we have introduced a new deep learning model to identify bad exposures in large imaging surveys.
Our model is implemented based on a pre-trained vision transformer in the context of the DINO framework.
We have explored the possibility of transferring information learned from a model trained on the ImageNet collection of general, feature-rich images to the classification of astronomical images from the LS. It is worth noting that this part of our work does not require any individual labels to infer patterns that correspond to different types of astronomical exposures.

We eventually classify our exposures based on metric distances in a high-dimensional space using a kNN approach. The training of the kNN {\em does} require a set of labeled bad exposures, for which we adopted a multi-class classification approach with 12 categories. We employed 11 categories for bad exposures, each describing a specific issue, and 1 category for good exposures. We compiled a bad exposure dataset using exposures flagged by human experts in LS and as part of the DELVE campaign. We treated exposures in LS DR9 that are not in the bad exposure dataset as generically \textit{good}, and drew exposures from this set to form the ``good" category. 

We ran the pretrained ViT to generate embeddings that were later used to train the kNN classifier. The performance of our ViT+kNN pipeline was evaluated in two steps.
We first evaluated the structure of the embeddings via clustering analysis to understand whether the model can successfully recognize and separate features that correspond to bad exposures.
We then evaluated the classification performance of the kNN using several metrics, including accuracy and F1 score.
Our model achieved a good and consistent performance with $>80\%$ classification precision for most categories of bad exposures.

We applied our pipeline to identify bad exposures in preparation for DR11 of the LS. Our model identified 946 bad exposures in DR11. Of these, 
159 bad exposures were already included in older exposures from DR9 and DR10 that will be reprocessed during DR11, and 786 bad exposures were identified in new exposures that will be processed for DR11.
When comparing our bad exposure list with the list generated by the {\tt legacypipe} software that is used to process LS images, we found 708 mutual ``bad'' exposures and 72 bad exposures that were only identified by our pipeline. Our newly identified bad exposures will be included in the list of exposures that will be rejected by {\tt legacypipe} when processing DR11 of the LS. These exposures are mostly in the \gs~category, which indicates that these exposures contain pupil ghosts and/or scattered light.
% before \lsp~processing stage.

With future large imaging survey telescopes, such as the Rubin Observatory \citep{LSSTDesign}, coming online, it will become increasingly impractical for human experts to identify bad exposures.
Our model is consistent with a new approach; applying self-supervised learning to efficiently identify bad exposures in large imaging surveys with minimal human intervention.
Moving forwards, new models and architectures, such as the Hierarchical Image Pyramid Transformer, may be of even more promise in both improving the classification accuracy for, and identifying various kinds of, bad exposures. Given the relative speed and scalability of our classification pipeline as described in Section \ref{sec:results},
it should also be possible to identify bad exposures on-the-fly, with the potential to adjust observing strategies in real-time.

The approach we adopted in this work represents a novel intersection of transfer learning and self-supervised learning in astronomical image analysis. By leveraging the power of pretrained self-supervised models, our classification of bad exposures benefited from both the sophisticated feature-learning capabilities of SSL and the rich representational power acquired through training on diverse datasets. In some sense, our model's ability to recognize patterns, textures, and structural anomalies transcends the domain gap between natural images and astronomical observations. In general, SSL may be a good choice for image classification in astronomy (and other domains) because it can provide a bridge between general computer vision and specialized scientific applications.

% \software{\code{astropy} \citep{astropy:2013, astropy:2018, astropy:2022}, \code{PyTorch} \citep{pytorch, Ansel_PyTorch_2_Faster_2024}, \code{scikit-learn} \citep{scikit-learn}, \code{matplotlib} \citep{matplotlib}}.

%% Also note that the acknowledgment environment does not support long amounts of text. If you have a lot of people and institutions to acknowledge, do not use this command. Instead, create a new \section{Acknowledgments}.
% \begin{acknowledgments}
\section*{Acknowledgements}

YL\ and ADM\ were supported by the U.S.\ Department of
Energy, Office of Science, Office of High Energy Physics,
under Award Number DE-SC0019022. YL was also partially supported by the University of Wyoming School of Computing Graduate Computing Scholarship and the Argonne National Labratory W.J.\ Cody Associate Program.
The work of AD and FV is supported by NOIRLab, which is managed by the Association of Universities for Research in Astronomy (AURA) under a cooperative agreement with the U.S. National Science Foundation.

The Legacy Surveys consist of three individual and complementary projects: the Dark Energy Camera Legacy Survey (DECaLS; Proposal ID \#2014B-0404; PIs: David Schlegel and Arjun Dey), the Beijing-Arizona Sky Survey (BASS; NOAO Prop. ID \#2015A-0801; PIs: Zhou Xu and Xiaohui Fan), and the Mayall z-band Legacy Survey (MzLS; Prop. ID \#2016A-0453; PI: Arjun Dey). DECaLS, BASS and MzLS together include data obtained, respectively, at the Blanco telescope, Cerro Tololo Inter-American Observatory, NSF’s NOIRLab; the Bok telescope, Steward Observatory, University of Arizona; and the Mayall telescope, Kitt Peak National Observatory, NOIRLab. Pipeline processing and analyses of the data were supported by NOIRLab and the Lawrence Berkeley National Laboratory (LBNL). The Legacy Surveys project is honored to be permitted to conduct astronomical research on Iolkam Du’ag (Kitt Peak), a mountain with particular significance to the Tohono O’odham Nation.
NOIRLab is operated by the Association of Universities for Research in Astronomy (AURA) under a cooperative agreement with the National Science Foundation. LBNL is managed by the Regents of the University of California under contract to the U.S. Department of Energy.

This project used data obtained with the Dark Energy Camera (DECam), which was constructed by the Dark Energy Survey (DES) collaboration. Funding for the DES Projects has been provided by the U.S. Department of Energy, the U.S. National Science Foundation, the Ministry of Science and Education of Spain, the Science and Technology Facilities Council of the United Kingdom, the Higher Education Funding Council for England, the National Center for Supercomputing Applications at the University of Illinois at Urbana-Champaign, the Kavli Institute of Cosmological Physics at the University of Chicago, Center for Cosmology and Astro-Particle Physics at the Ohio State University, the Mitchell Institute for Fundamental Physics and Astronomy at Texas A\&M University, Financiadora de Estudos e Projetos, Fundacao Carlos Chagas Filho de Amparo, Financiadora de Estudos e Projetos, Fundacao Carlos Chagas Filho de Amparo a Pesquisa do Estado do Rio de Janeiro, Conselho Nacional de Desenvolvimento Cientifico e Tecnologico and the Ministerio da Ciencia, Tecnologia e Inovacao, the Deutsche Forschungsgemeinschaft and the Collaborating Institutions in the Dark Energy Survey. The Collaborating Institutions are Argonne National Laboratory, the University of California at Santa Cruz, the University of Cambridge, Centro de Investigaciones Energeticas, Medioambientales y Tecnologicas-Madrid, the University of Chicago, University College London, the DES-Brazil Consortium, the University of Edinburgh, the Eidgenossische Technische Hochschule (ETH) Zurich, Fermi National Accelerator Laboratory, the University of Illinois at Urbana-Champaign, the Institut de Ciencies de l’Espai (IEEC/CSIC), the Institut de Fisica d’Altes Energies, Lawrence Berkeley National Laboratory, the Ludwig Maximilians Universitat Munchen and the associated Excellence Cluster Universe, the University of Michigan, NSF’s NOIRLab, the University of Nottingham, the Ohio State University, the University of Pennsylvania, the University of Portsmouth, SLAC National Accelerator Laboratory, Stanford University, the University of Sussex, and Texas A\&M University.

BASS is a key project of the Telescope Access Program (TAP), which has been funded by the National Astronomical Observatories of China, the Chinese Academy of Sciences (the Strategic Priority Research Program “The Emergence of Cosmological Structures” Grant \# XDB09000000), and the Special Fund for Astronomy from the Ministry of Finance. The BASS is also supported by the External Cooperation Program of Chinese Academy of Sciences (Grant \# 114A11KYSB20160057), and Chinese National Natural Science Foundation (Grant \# 12120101003, \# 11433005).

The Legacy Survey team makes use of data products from the Near-Earth Object Wide-field Infrared Survey Explorer (NEOWISE), which is a project of the Jet Propulsion Laboratory/California Institute of Technology. NEOWISE is funded by the National Aeronautics and Space Administration.

The Legacy Surveys imaging of the DESI footprint is supported by the Director, Office of Science, Office of High Energy Physics of the U.S. Department of Energy under Contract No. DE-AC02-05CH1123, by the National Energy Research Scientific Computing Center, a DOE Office of Science User Facility under the same contract; and by the U.S. National Science Foundation, Division of Astronomical Sciences under Contract No. AST-0950945 to NOAO.

\appendix

\section{Final Training and Testing Samples}

Our final training and testing dataset comprises two parts: the combined bad exposure list and the sample of good exposures.
The final bad image list incorporates the DECam bad exposure list and the human-expert lists from DES and DELVE.
%We selected all CCD images from the exposures listed in the DECam bad exposure list.
The total number of labeled bad exposures is 73{,}990, and detailed numbers from each source in each category are shown in Table \ref{tab:category}.

As described in Section \ref{subsec:legacypipeFlag}, good exposure candidates are all exposures not identified in the combined bad image set or flagged by {\tt legacypipe}. We selected 10{,}000 good exposures from the candidate set to balance the training dataset, and the final number of training exposures is 40{,}181.  

\begin{table}[htbp]
    \centering
    \begin{tabular}{|c|c|c|c|c|c|}
        \hline
        Bad Category & DECaLS list & DES + DELVE & from both & Total\\
        \hline
        \textit{saturated} & 778 & 0 & 0 & 778 \\
        \textit{clouds or bad transparency} & 992 & 22331 & 15 & 23338 \\
        \textit{psf} & 2359 & 4781 & 501 & 7641 \\
        \textit{non-optimal exposure} & 3194 & 0 & 0 & 3194 \\
        \textit{Ghost/Scatter} & 0 & 15755 & 1 & 15756 \\
        \textit{NObjects} & 0 & 10093 & 0 & 10093 \\
        \textit{Bad CCD}& 0 & 2621 & 2 & 2623 \\
        \textit{Noise} & 0 & 2406  & 61 & 2467 \\
        \textit{Wonky} & 0 & 667 & 0 & 667 \\
        \textit{Telescope Moving} & 725 & 1403 & 601 & 2729 \\
        \textit{out of focus} & 2160 & 0 & 180 & 2340 \\
        \hline
        Total & 10208 & 60057 & 1361 & 71626 \\
        \hline
    \end{tabular}
    \caption{The categories of the labeled bad exposures with the number of CCD images in each category from each  source. The ``DECaLS list'' and ``DES+DELVE'' columns include data exclusively from those sources. The ``from both'' column counts the exposures that are identified in both lists. As discussed in Sec.\ \ref{subsec:badExpList} and \ref{subsec:expertLabel}, not all categories appear in both samples. Images counted in this table are selected to avoid dust extinction, as discussed in Sec.\ \ref{subsec:extCut}.}
    \label{tab:category}
\end{table}

\newpage

\section{Sample Exposures Identified by the Model}\label{app:sample-exp}

\begin{figure}
    \centering
    \gridline{
    \fig{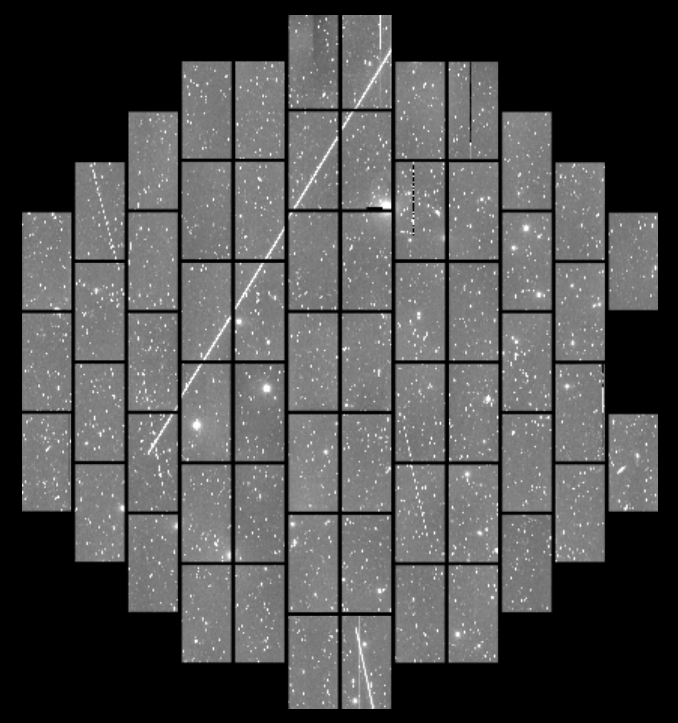}{0.35\linewidth}{(a1) expnum=246861}
    \fig{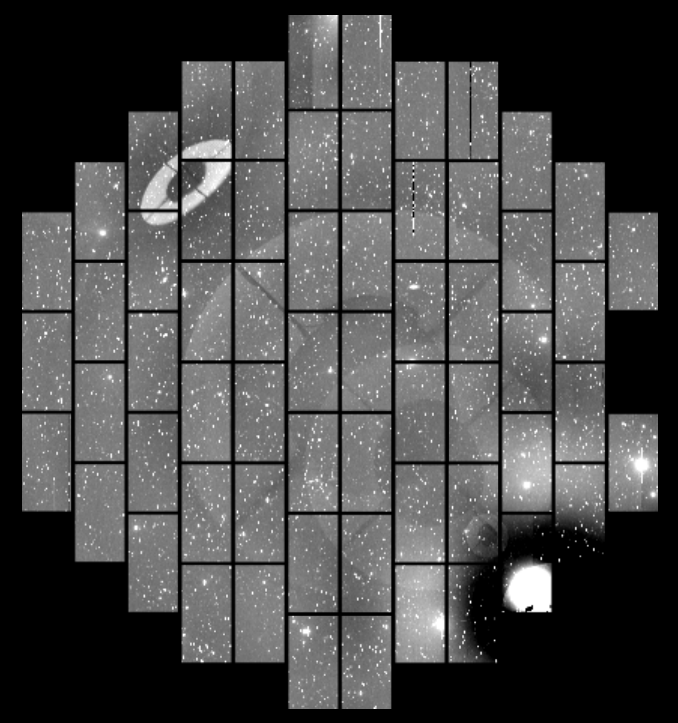}{0.35\linewidth}{(b1) expnum=624625}
    }
    \gridline{
    \fig{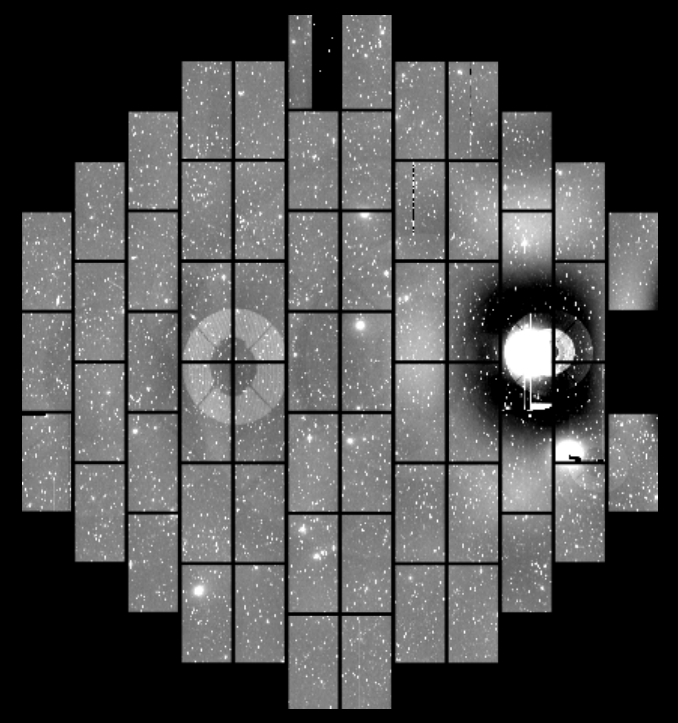}{0.35\linewidth}{(a2) expnum=635208}
    \fig{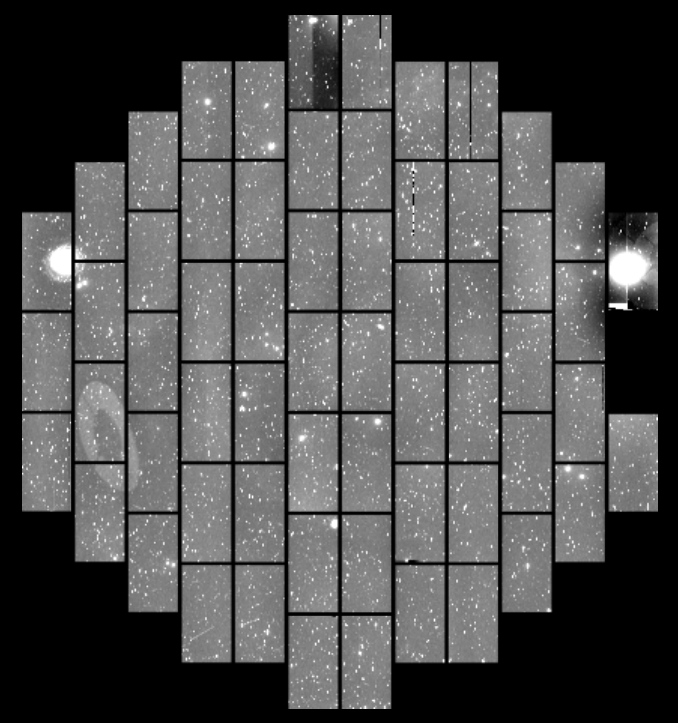}{0.35\linewidth}{(b2) expnum=730449}
    }
    \gridline{
    \fig{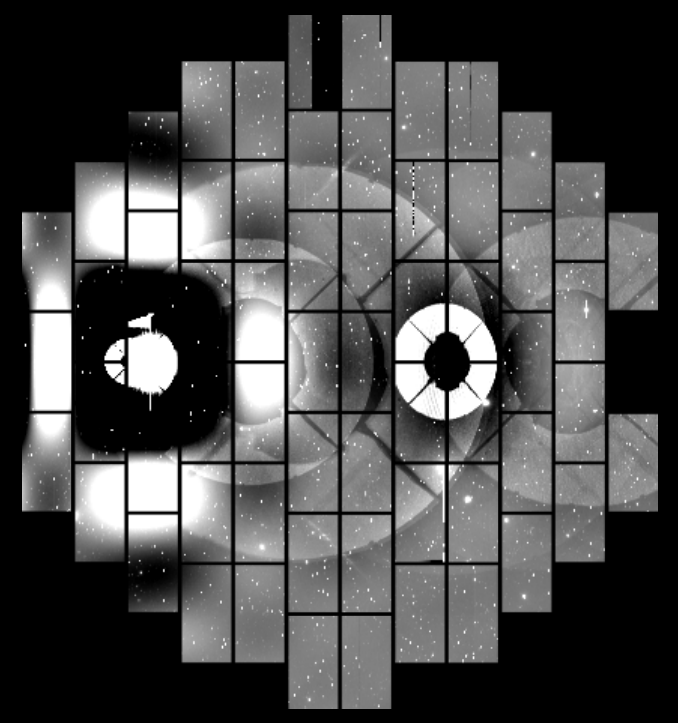}{0.35\linewidth}{(a3) expnum=697269}
    \fig{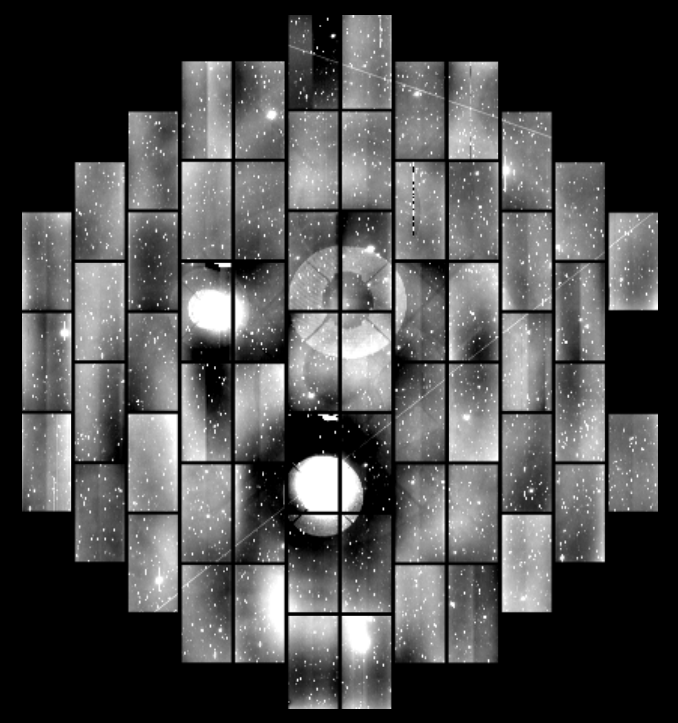}{0.35\linewidth}{(b3) expnum=862944}
    }
    
    \caption{A set of sample exposures drawn from the two clusters depicted in Figure \ref{fig:tsne-3D}. The left-hand (right-hand) set of three panels show exposures that correspond to the red (green) cluster in Figure \ref{fig:tsne-3D}.}
    \label{fig:sample-exp}
\end{figure}

\begin{figure}
    \centering
    \gridline{
        \fig{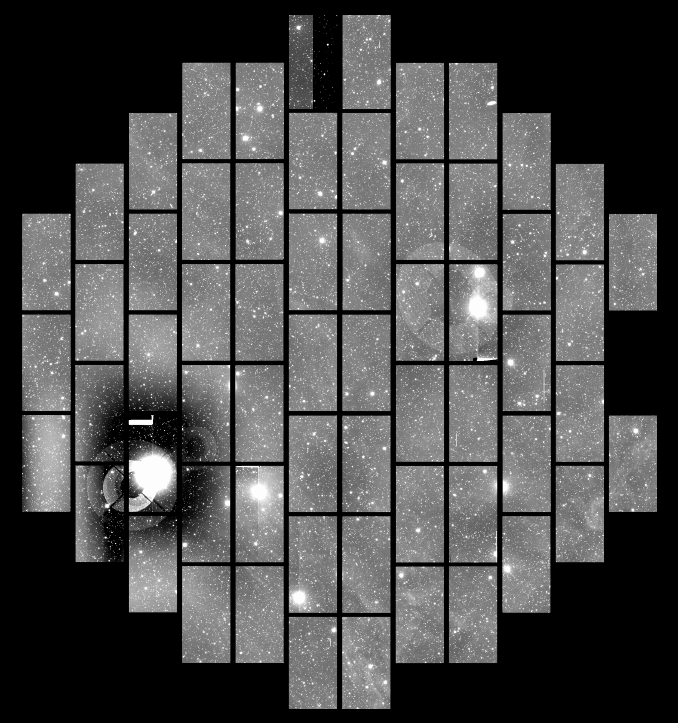}{0.5\textwidth}{(a) Exposure 1222425 (r-band)}
        \fig{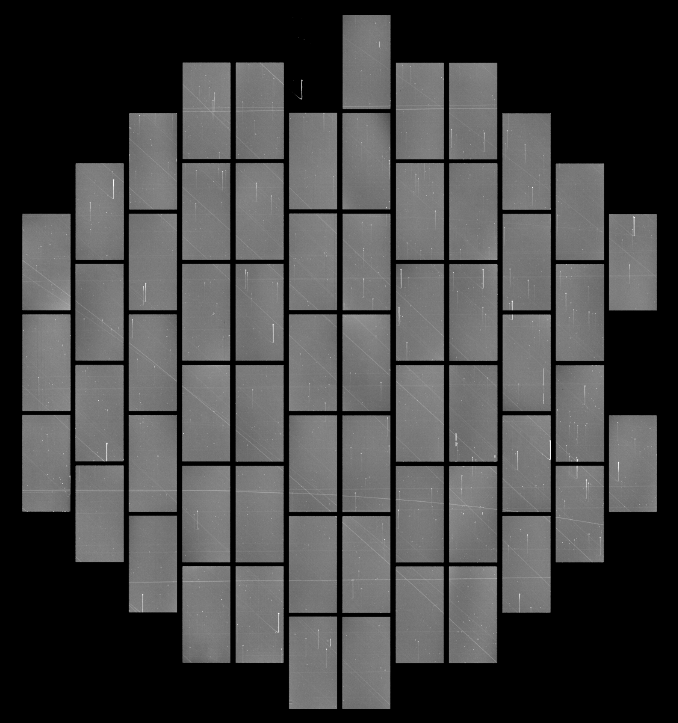}{0.5\textwidth}{(b) Exposure 1217078 (i-band)}
    }
    \caption{Two exposures in DR11 that are classified as ``bad'' by our pipeline. Our ML labels the exposure on the left (right) as \textit{Ghost\_Scatter} (\textit{Telescope\_Moving}). These classifications have subsequently been verified by human experts.}
    \label{fig:inference-success}
\end{figure}

\begin{figure}
    \centering
    \includegraphics[width=0.5\linewidth]{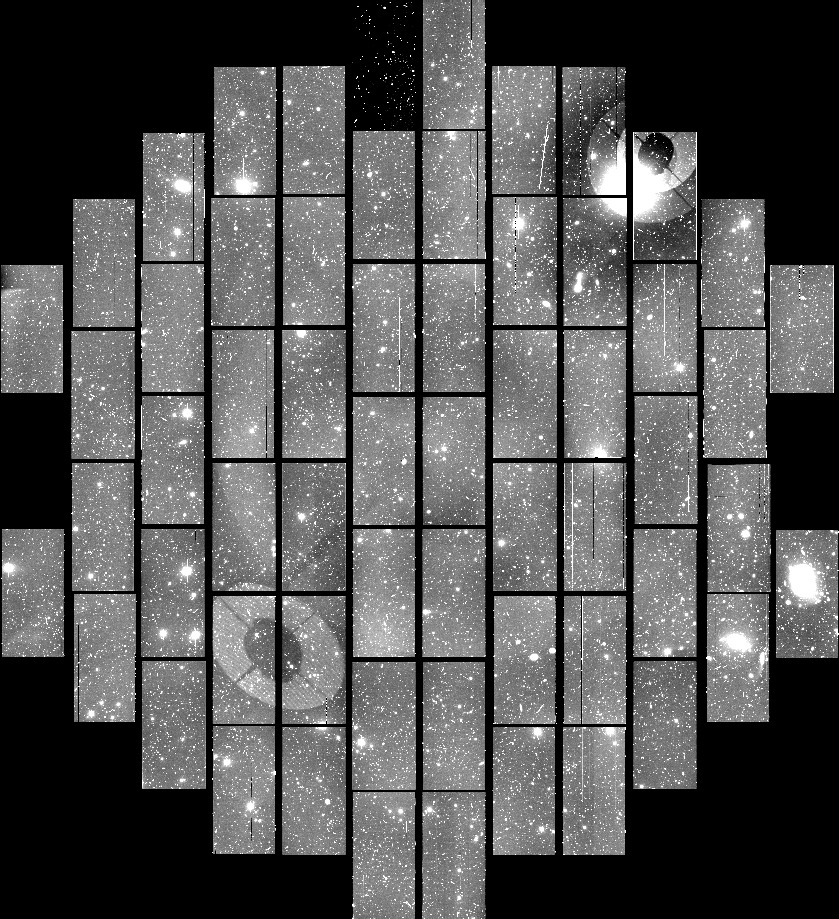}
    \caption{Exposure 260867 from the DECaLS ($g$-band). %This 90-second exposure was taken on mjd=56632.177. 
    The LS DR9 included this exposure, while our pipeline flagged it as bad in the \gs\ category. Human experts later confirmed that this exposure was bad during visual inspection. Several pupil ghosts are present in this exposure, which are likely caused by the bright star in the top-right.}
    \label{fig:bad-exp-ghost}
\end{figure}

\begin{figure}
    \centering
    \includegraphics[width=0.5\linewidth]{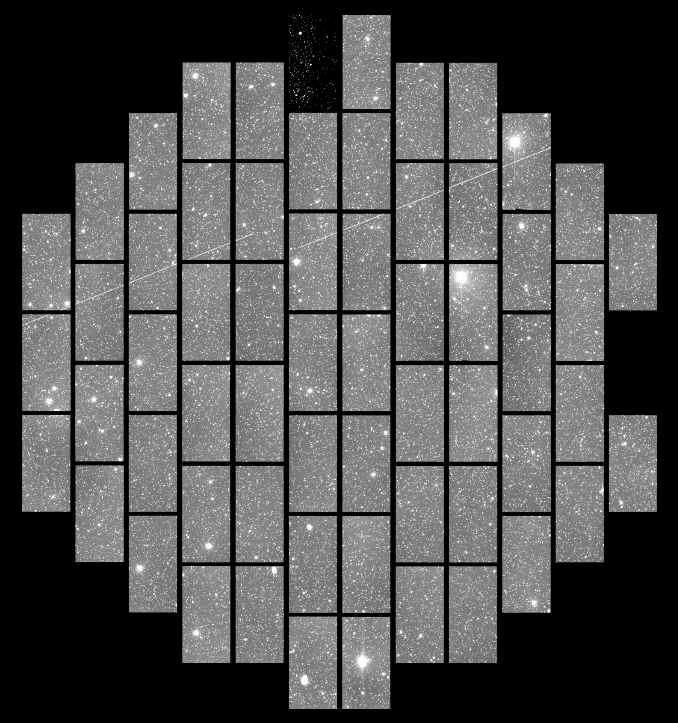}
    \caption{Exposure 1302430 ($i$-band). This is an example of an image that was misclassified as ``bad'' (in the \tm\ category) by our pipeline. Human experts later evaluated this exposure as good during visual inspection. The misclassification is likely due to the two parallel and proximate satellite trails, which are a novel feature that might mimic movement of the telescope.}
    \label{fig:mis-tm}
\end{figure}

\newpage

\section{Data Availability Statement}\label{app:data}

The DECaLS DR9 and DR10 data are available at \url{https://www.legacysurvey.org/}. The DR11 data is expected to be available later in 2025. The DECam bad exposure list, as compiled by LS for LS DR10.2, is available at \url{https://github.com/legacysurvey/legacypipe/blob/DR10.2/py/legacyzpts/data/decam-bad_expid.txt}. The DES and DELVE labeled bad exposures are available at \url{https://github.com/delve-survey/exclude}.
All CP-reduced data for DECam are available from NOIRLab at \url{https://astroarchive.noirlab.edu/}.

\section{Reproducibility Considerations of ML models}

We follow a number of best practices to make our result as reproducible as possible \citep[see, e.g.,][]{Luo_2024}. 
For example, our source code and analysis notebooks are available in a public repository \url{https://github.com/Brookluo/ssl-bad-exposure-identification/releases/tag/v1.0}, and we have made sure to fix potentially changeable quantities, such as random seeds, software versions, and runtime environments. However, even following best practices to guard against irreproducibility, we recognize that our results may not be fully ``bit-wise'' reproducible. This is because the ML model we employ is mostly trained stochastically, so the results may vary slightly between different runs of our pipeline. But, we expect any subsequent results to be {\em consistent} with the reported results in this paper provided the pretrained ViT model remains unchanged on PyTorch Hub.


\bibliography{main}{}
\bibliographystyle{aasjournal}



\end{document}